\documentclass{JHEP3}

\usepackage[all]{xy}
\usepackage{amsmath,amssymb}
\usepackage{graphicx}

\usepackage{epstopdf}

\def\cA{{\cal A}}

\def\cC{{\cal C}}
\def\cD{{\cal D}}
\def\cE{{\cal E}}

\def\cH{{\cal H}}

\def\cM{{\cal M}}

\def\cR{{\cal R}}
\def\cS{{\cal S}}

\def\dZ{{\Bbb Z}}

\def\dR{{\Bbb R}}

\def\ot{{\otimes}}

\def\Cech{\u{C}ech}
\def\d{\mbox{d}}

\def\be{\begin{equation}}
\def\ee{\end{equation}}
\def\beq{\begin{eqnarray}}
\def\eeq{\end{eqnarray}}

\def\lp{ \left(}
\def\rp{ \right)}
\def\la{ \left\{}
\def\ra{ \right\}}
\def\lb{ \left[}
\def\rb{ \right]}

\newcommand{\ii}{\mathrm{i}}

\title{Quasi-quantum groups from Kalb-Ramond fields and magnetic amplitudes for strings on orbifolds}

\author{Jan-H. Jureit\\ Christian-Albrechts-Universit\"at zu Kiel\\
24098 Kiel \\ GERMANY \\ E-mail: \email{jureit@cpt.univ-mrs.fr}}
\author{Thomas Krajewski\\ Centre de Physique
Th\'eorique\footnote{Unit\'e Mixte de Recherche (UMR) 6207 du CNRS
et des Universit\'es Aix-Marseille 1 et 2 \\ Sud Toulon-Var,
Laboratoire affili\'e \`a la
FRUMAM (FR 2291)} \\ CNRS--Luminy, Case 907\\ 13288 Marseille Cedex 9\\
FRANCE \\ E-mail: \email{krajew@cpt.univ-mrs.fr}}

\abstract{We present the general form of the operators that lift the group
action on the twisted sectors of a bosonic string on an
orbifold ${\cal M}/G$, in the presence of a Kalb-Ramond field strength
$H$. These operators turn out
to generate the quasi-quantum group $D_{\omega}[G]$, introduced in the
context of orbifold conformal field theory by R. Dijkgraaf, V. Pasquier and
P. Roche. The 3-cocycle  $\omega$ entering in the definition of
$D_{\omega}[G]$ is related to $H$ by  a series of cohomological
equations in a tricomplex combining de Rham, \Cech{} and group coboundaries. We
construct magnetic amplitudes for the twisted sectors and show that
$\omega=1$ arises as a consistency condition for the orbifold
theory. Finally, we recover discrete torsion as an ambiguity in the
lift of the group action to twisted sectors, in accordance with
previous results presented by E. Sharpe.}

\preprint{hep-th/1234567}

\begin{document}

\section{Introduction}
Over the last decade, important breakthroughs occurred in the theory of
strings in magnetic backgrounds \cite{magnetic}. Most of
these developments were triggered by the identification of
noncommutative geometry \cite{connes} as a natural framework for an effective
description of the dynamics in such backgrounds.  Roughly speaking,
geometric properties of fields, strings and branes are formulated in
terms of noncommutative algebras that are deformations, induced by
a magnetic background, of otherwise commutative algebras.

The most celebrated example of such a noncommutative algebra is the
noncommutative torus, obtained by inserting phase factors in the
commutation rules of the Fourier modes. From an elementary
physics viewpoint, the noncommutative torus can be thought of as a
discrete magnetic translation algebra. More generally, one can
consider the motion of a particle in an external
$B$-field invariant under a group $G$. Then, it is  a general result
that the operators $T_{g}$ that lift the action of $G$ to the wave
functions only form a
projective representation of the group $G$,
$T_{g}T_{h}=\omega_{g,h}\,T_{gh}$ where $\omega$ is a group
2-cocycle.

In this paper, we present a generalization of this construction
pertaining to bosonic strings on the orbifold ${\cal M}/G$, in the presence
of a Kalb-Ramond field strength $H$ on ${\cal M}$, invariant under
$G$. In analogy with the case of a particle in a $B$-field, we define
the {\it stringy magnetic translations} as the operators $T_{g}^{w}$
that realize the action
of $g\in G$ on the twisted sectors made of strings of winding $w$. The
general form of these operators follows from their commutation with
propagation along cylinders,
\begin{equation}
T\quad\parbox{1.5cm}{\mbox{\includegraphics[width=1.5cm]{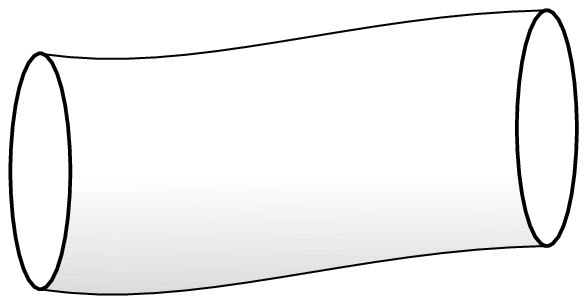}}}
\quad=\quad
\parbox{1.5cm}{\mbox{\includegraphics[width=1.5cm]{./pics/_minicy.eps}}}
\quad T.
\end{equation}
Because string interactions are by essence geometrical, the
algebra generated by these operators admits a richer
structure than in the case of a particle. Indeed,  the action of
$T_{g}^{w}$ on two string states is constrained by its commutation with
processes involving pairs of pants, which provides the algebra with a
coproduct $\Delta$,
\begin{equation}
T\quad
\parbox{1.5cm}{\mbox{\includegraphics[width=1.5cm]{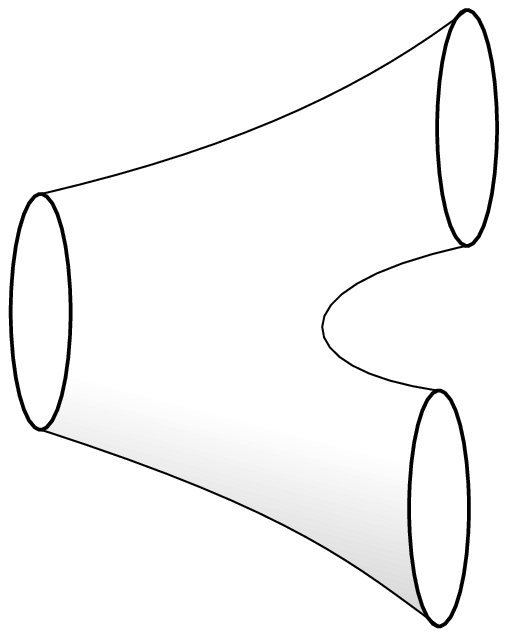}}}
\quad=\quad
\parbox{1.5cm}{\mbox{\includegraphics[width=1.5cm]{./pics/_minipant.eps}}}
\quad\Delta T.
\end{equation}
As a result, the stringy magnetic translations turn out to generate the
{\it quasi-quantum group} $D_{\omega}[G]$ introduced by Dijkgraaf,
Pasquier and Roche in the context of orbifold conformal field theory
\cite{dijkgraaf}, where $\omega$ is a 3-cocycle derived from an
analysis of the invariance properties of the potentials of the
field strength $H$.

To construct the orbifold string theory, some care is needed when
dealing with the Kalb-Ramond field. Indeed, whereas the field strength
$H$ is invariant, this is not necessarily so for its
potential $B$. Moreover, in many interesting cases the relation $H=dB$
only holds locally on ${\cal M}$, so that one has to cover ${\cal M}$
by patches and work with locally defined potentials. In order to handle all
these fields and their transformation laws under $G$,
it is convenient to introduce a tricomplex combining de Rham, \Cech{} and group
cohomologies. Starting with $H$, we obtain a sequence of locally
defined fields by solving cohomological equations in the tricomplex. The
equations relating these fields are equivalent to those derived by
E. Sharpe in his analysis of discrete torsion \cite{sharpe}, where the
action of $G$ has been lifted to the gauge fields. The construction of
the operators $T_{g}^{w}$ is a natural continuation of his work, since
they allow to further lift the group action to the string's wave function.

The locally defined potentials are the basic ingredients entering in
the construction of the magnetic amplitudes for the twisted
sectors. The latter are the
phases describing the coupling of a string to the Kalb-Ramond
field that have to be inserted in the world-sheet path integral. They are
built using a minor modification of the closed string magnetic
amplitudes constructed by K. Gaw\c{e}dzki \cite{gawedzki}, in order to deal
with the twisted sectors. We construct in detail the
amplitudes for the cylinder and the pair of pants in order to derive the
operators $T_{g}^{w}$ and their coproduct. Whereas this construction is
valid for arbitrary $\omega$, general magnetic amplitudes are flawed
by global anomalies unless $\omega$ is trivial. This is similar to
what happens for a particle: Wave functions on the quotient ${\cal M}/G$
are well defined only if the 2-cocycle $\omega$ is trivial.

This paper is organized as follows.

Section 2 is devoted to a detailed account of the particle's case. This
serves to illustrate on a very simple example the reasoning we shall
employ for strings in later sections. In particular, we derive the
magnetic translations $T_{g}$ both in the canonical and  path
integral formalism.

In section 3, we introduce the tricomplex and solve the cohomological
equations for a particle and a string. This yields the fields
needed in the sequel, as well as their gauge transformations. Since this
technique is by no means restricted to the Kalb-Ramond
field, we illustrate it briefly on the M-theory 3-form.

In Section 4, we present the magnetic amplitudes for a cylinder
and  construct the stringy magnetic translations $T_{g}^{w}$. Then,
we show that they obey the multiplication law of the quasi-quantum
group $D_{\omega}[G]$.

Section 5 deals with interacting strings. We first construct the
amplitude for the pair of pants, from which we derive the coproduct of
$D_{\omega}[G]$. Then, we illustrate the Drinfel'd associator and the
braiding from the point of view of tree level string
interactions. Besides, we discuss loop amplitudes and illustrate the
global anomalies for the torus in the case $\omega\neq1$. Finally, we
show how discrete torsion arises in our context.

An appendix gathers some useful facts about quasi Hopf algebras, as
can be found, for instance, in \cite{kassel}.

\section{Particle in a magnetic field}
\label{particle}
\subsection{Magnetic amplitude}

To gain some insight into the string case, it is helpful to first
investigate the case of a particle moving on a manifold ${\cal M}$ in
the presence of a magnetic field, which is a closed 2-form $B$.

The $B$-field  needs not to be globally exact so that  we have to cover
${\cal M}$ by a good open cover $\la U_{i}\ra$ made out of contractible open
sets. This always exists if we assume that ${\cal M}$ is paracompact,
as is the case for a finite dimensional manifold. On the chart $U_{i}$
there exists a real valued 1-form $A_{i}$ such that $\d A_{i}=B_{i}$,
where $B_{i}$ is the restriction of $B$ to $U_{i}$. On the
intersection $U_{i}\cap U_{j}$,
there exists  a $U(1)$-valued function $f_{ij}$, with
$f_{ij}=(f_{ji})^{-1}$ such that $\mathrm{i}\,\d\log
f_{ij}=A_{j}-A_{i}$, because the latter is a closed
1-form. Finally, on triple intersections $U_{i}\cap U_{j}\cap U_{k}$ one
requires that $f_{ij}f_{jk}f_{ki}=1$. This last condition imposes that
the de Rham cohomology class of $B$ belongs to
$\mathrm{H}^{2}({\cal M},2\pi\dZ)$,
which is equivalent to stating that its integral over any closed
surface in ${\cal M}$ lies in $2\pi\dZ$.

From a geometrical viewpoint, this means that $f_{ij}$ are the
transition functions of a hermitian line bundle $\cE$ over ${\cal
  M}$, whose sections are defined locally on $U_{i}$ by complex valued
functions $\psi_{i}$, subjected to $\psi_{i}=f_{ij}\psi_{j}$ on double
intersections. The de Rham
differential $\d$ and the fields $A_{i}$ define a connection on this
bundle by $(\nabla\psi)_{i}=\d\psi_{i}-\mathrm{i}A_{i}\psi_{i}$, with
globally defined curvature $B$. Besides, this bundle is equipped with a natural
hermitian form taking two sections $\psi$ and $\chi$ to the globally
defined function $\psi^{\ast}_{i}\chi_{i}$. In physics, sections of
$\cE$ are the wave functions of the particle, and equipped with a
natural scalar product they provide
the Hilbert space of states $\cH$. The connection is the covariant
derivative that results from the minimal coupling prescription.

Given a magnetic field $B$, $A_{i}$ and $f_{ij}$ are not
unique. First, one can replace $A_{i}$ by
$A_{i}+\mathrm{i}\,\mathrm{d}\log \eta_{i}$, which induces the replacement of
$f_{ij}$ by $(\eta_{i})^{-1}f_{ij}\eta_{j}$ and of  $\psi_{i}$ by
$(\eta_{i})^{-1}\psi_{i}$.
The bundle and the connection are traded for equivalent ones, and this
is just a gauge transformation. If the transition functions are
unchanged, this can be interpreted as a gauge transformation within
the same bundle, as is more common in physics.
In what follows, we always use the expression "gauge transformation"
in the general sense of an equivalence of bundles with connection.

There is a second ambiguity in the
definition of $f_{ij}$ alone, that can be replaced by $\alpha_{ij}f_{ij}$,
with constant $\alpha_{ij}\in U(1)$. Consistency conditions on triple
intersections require that $\alpha_{ij}\alpha_{jk}\alpha_{ki}=1$,
which means that $\alpha_{ij}$ defines a constant \Cech{} cocycle. Up to
constant gauge transformations, those ambiguities are classified by a
set of angles in $\mathrm{H}^{1}({\cal M},U(1))$, that represent inequivalent
quantizations of the same classical theory. These angles label the
various sectors of the theory and are often referred to as {\it vacuum
  angles}. Whereas gauge ambiguities are always present, vacuum angles
exist only if the cohomology group $\mathrm{H}^{1}({\cal M},U(1))$ is non
trivial.
\medskip

A fundamental object in quantum physics is the propagator $K(y,x)$
which is the kernel of the evolution operator. In quantum mechanics,
it is the transition amplitude from $x$ to $y$  and it
provides the time evolution of the wave function $\psi\rightarrow\psi'$ as
\be
\psi'(y)=\int dx\, K(y,x)\psi(x).\label{evolution}
\ee
Ignoring its distributional nature, $K(y,x)$ can be thought of as a
linear map from the fiber at $x$ of $\cE$ to its fiber at $y$.

In Feynman's path integral approach, the propagator is obtained by summing over
all paths joining $x$ to $y$ an amplitude associated to each of these
paths. A path is a map $\varphi$ from a fixed interval $[a,b]$ into
${\cal M}$ and the euclidian propagator reads, in the absence of a magnetic
field,
\begin{equation}
K(y,x)=\mathop{\int[D\varphi]}
\limits_{\begin{subarray}{c}\varphi(a)=x\cr\varphi(b)=y\end{subarray}}\,
\mathrm{e}^{-S[\varphi]},\label{feynmanpart}
\end{equation}
where $S[\varphi]$ is the euclidian action. The integrand of
\eqref{feynmanpart} is referred to as the euclidian amplitude of the path.

In this setting, the effect of the $B$-field is encoded in the
magnetic amplitude that multiplies the euclidian amplitude,
\begin{equation}
\mathrm{e}^{-S[\varphi]}\xrightarrow{B-\mathrm{field}}
\mathrm{e}^{-S[\varphi]}{\cal A}[\varphi].
\end{equation}
 The magnetic amplitude ${\cal A}[\varphi]$ is constructed using the
 local data $(A_{i},f_{ij})$ as follows
\cite{gawedzki}, with the sign convention of  \cite{newgawedzki}. For
each $\varphi$, choose a triangulation $(l,v)$ of
the interval into segments $l_{\alpha}$ such that $\varphi(l_{\alpha})$
is entirely contained in an open set $U_{i_{\alpha}}$ and vertices
$v_{\beta}$ that bound the segments. For any vertex $v_{\beta}$ of the
triangulation, also choose an open set
$U_{j_{\beta}}$ such that $\varphi(v_{\beta})\in U_{j_{\beta}}$. The
endpoints $a$ and $b$ are included as vertices of the triangulation
and the open sets used to cover the corresponding points $x$ and $y$
are labelled by $i$ and $j$.
%
\begin{figure}[h]\centering
\includegraphics[width=8cm]{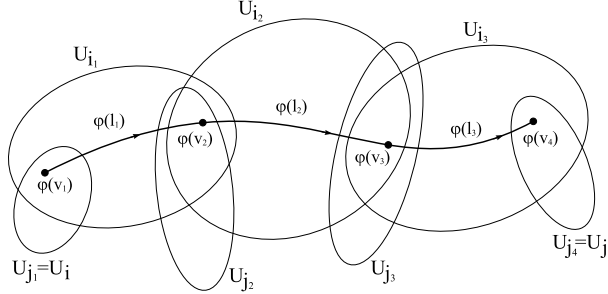}
\caption{Covering of a path with open sets.}
\label{_path}
\end{figure}
%

The magnetic amplitude of the path is defined as
\be
%
\cA_{ji}[\varphi]=\left[\exp \mathrm{i}\la\sum_{l_{\alpha}\in l}\int_{l_{\alpha}}
\varphi^{\ast}A_{i_{\alpha}}\ra\,
\mathop{\prod}_{\begin{subarray}{c} l_{\alpha}\in l\\ v_{\beta}\in \partial l_{\alpha}
\end{subarray}}
f_{i_{\alpha}j_{\beta}}^{-\epsilon_{\alpha\beta}}(\varphi(v_{\beta}))\right]_{ji},
\label{holo}
\ee
where $\epsilon_{\alpha\beta}=+1$ if $l_{\alpha}$ is arriving at
$v_{\beta}$ and $-1$ if it is leaving. It is easy
to check that this amplitude is independent of the choice of the
triangulation and of the assignment of an open set to each segment
 and inner vertex, but it depends  on the open sets $U_{i}$ and
 $U_{j}$ used to cover the endpoints $x$ and $y$. If we trade the latter for
$U_{k}$ and $U_{l}$, then the amplitude becomes
\be
\cA_{lk}[\varphi]=f_{lj}(y)\,\cA_{ji}[\varphi]f_{ik}(x),
\ee
in agreement with its interpretation as a map from the fiber at $x$ to
that at $y$. It is also invariant under
reparametrizations of the interval, since they only induce a change of
the triangulation. In terms of the complex valued functions $\psi_{i}$
and $\psi'_{i}$ defining the sections $\psi$ and $\psi'$,
\eqref{evolution} reads,
\begin{equation}
\psi'_{j}(y)=\sum_{i}\int dx\,\rho_{i}(x)\,K_{ji}(y,x)\,\psi_{i}(x),
\label{localK}
\end{equation}
where $\rho_{i}$ is a partition of unity subordinate to the cover
$U_{i}$. The kernel $K_{ji}(y,x)$ is computed using Feynman's
path integral as
\begin{equation}
K_{ji}(y,x)=
\mathop{\int[D\varphi]}
\limits_{\begin{subarray}{c}\varphi(a)=x\cr\varphi(b)=y\end{subarray}}\,
\mathrm{e}^{-S[\varphi]}{\cal A}_{ji}[\varphi].\label{localpath}
\end{equation}
Note that $S[\varphi]$ only involves the kinetic term and does not require the use of the local
potentials $(A_{i},f_{ij})$.

From an invariant point of view, all these
statements are obvious since the magnetic amplitude is nothing but the
holonomy of the connection along the path defined by $\varphi$.

Let us also note that even if we have been working here in the
euclidian setting, the magnetic term takes the same form as in the
real-time formalism. In this last case, an electric term may be included as
the time component of a 2-form on the space-time manifold. Besides,
our construction does not depend on the precise form of $S[\varphi]$
and applies to relativistic particles as well as to non relativistic
ones, the magnetic amplitude being always given by \eqref{holo}. More
generally, one could consider
${\cal M}$ as the configuration space of an arbitrary system, in such
  a way that it applies directly to strings if we choose for ${\cal
    M}$ the space of all embeddings of the string in
  space-time. However, it is more convenient to take a different
  route, based on the world-sheet magnetic amplitude.

\subsection{Projective group action on particle states}

\label{particleproj}

Consider now a finite group $G$, acting on ${\cal M}$ from the right. We
choose a right action of the group on the manifold ${\cal M}$, i.e.
$(x\!\cdot\!
g)\!\cdot\! h=x\cdot (gh)$ for any $x\in{\cal M}$ and $g,h\in G$, so that
we have a standard left action of $G$ by pull-back on differential
forms. The pull-back action is defined for functions as
$g^{\ast}f(x)=f(x\!\cdot\!g)$ and more generally for differential forms by
$\int_{\cal N}g^{\ast}\omega=\int_{{\cal N}\!\cdot g}\,\omega$ for any
$n$-form $\omega$ and  $n$-dimensional submanifold ${\cal
  N}\subset{\cal M}$. The pull-back action is also compatible with the
de Rham differential in the sense that
$\d\left(g^{\ast}\omega\right)=g^{\ast}\left(\d\omega\right)$.
These properties of the pull-back action of $G$ are essential in the
following computations.

Let us now assume that the magnetic field $B$ considered in the
previous section  is invariant under the group
action, which translates into the equation $g^{\ast}B=B$ for any $g\in
G$. We aim at finding operators $T_{g}$ that lift the action of $G$ to the wave
functions of the particle in such a way that the dynamics remains
invariant. We assume that the action $S[\varphi]$ is genuinely invariant,
i.e. that $S[\varphi\!\cdot\!g]=S[\varphi]$ for any path $\varphi$ and
$g\in G$. This is the case for a particle moving on the euclidian
plane in the presence of an external uniform magnetic field, as we
shall see at the end of this section.

To implement the group action on the locally defined gauge fields, it
is convenient to work with a good invariant cover $\la U_{i}\ra$ of
${\cal M}$ by open sets $U_{i}$ that are stable under $G$ and which are
disjoint unions of contractible open sets. Acting with $G$ on a
section defined by $\psi_{i}$ on $U_{i}$, we find that $g^{\ast}\psi_{i}$ is a
section of the pullback bundle $g^{\ast}\cE$. The latter is defined by the
transition functions $g^{\ast}f_{ij}$ and is equipped with
the pullback connection given by the forms $g^{\ast}A_{i}$. Because $B$
is invariant under $G$, the two connections have the same curvature
$B$.  If  $\mathrm{H}^{1}({\cal M},U(1))$ is trivial, which is the case for
simply connected ${\cal M}$, the bundles $\cE$ and $g^{\ast}\cE$ are
isomorphic as bundles with connections,  so that there exists a $\phi_{g;\,i}$ such that
\be
\la
\begin{array}{rclrl}
g^{\ast}A_{i}-A_{i}&=&\ii\,\d\log\phi_{g;\, i} &\mbox{on}& U_{i},\\
g^{\ast}f_{ij}\,(f_{ij})^{-1}&=&\phi_{g;\,j}\,(\phi_{g;\,i})^{-1}
&\mbox{on}& U_{i}\cap U_{j}.
\end{array}
\right.
\label{phipart}
\ee
$\phi_{g;i}$ determines the isomorphism since it takes a section of
${\cal E}$ to a section of $g^{\ast}{\cal E}$ as
$\psi_{i}\mapsto(\phi_{g;i})^{-1}\psi_i$. Note that the explicit
construction of $\phi_{g;i}$ can be performed using \Cech{}
cohomology: One solves the first equation in \eqref{phipart},
then the triviality of the cohomology group $\mathrm{H}^{1}({\cal M},U(1))$
shows that there is a common solution to both equations, unique up to
a multiplicative constant.

Using this isomorphism, the unitary operators $T_{g}:\,\cH\rightarrow\cH$ are defined locally by
\be
(T_{g}\psi)_{i}=\phi_{g;\,i}\,g^{\ast}\psi_{i},\label{localT}
\ee
for any section $\psi$ defined by the complex valued functions $\psi_{i}$.
As is easy to check, $(T_{g}\psi)_{i}$ also defines a section of ${\cal
  E}$ and $T_{g}$ is composed of two operations. First, we take the
pull-back of $\psi$ which is a section of $g^{\ast}{\cal E}$ and then use
the inverse of the previous isomorphism to come back to a section of
${\cal E}$.
\medskip

The map $g\mapsto T_{g}$ provides a lift to the quantum setting of the
classical symmetry group $G$. By construction, these operators commute
with the connection $\nabla$ so that they also commute with the
hamiltonian that is constructed out of $\nabla$. As expected in
quantum mechanics, $g\mapsto T_{g}$ only provides a projective
representation of $G$,
\be
T_{g}T_{h}=\omega_{g,h}\,T_{gh}
\ee
with
\be
\omega_{g,h}=\frac{g^{\ast}\phi_{h;\,i}\,\phi_{g;\,i}}{\phi_{gh;\,i}}.
\label{2cocycle}
\ee
As is easy to check, $\omega$ is constant and does not depend on the
open set $U_{i}$ used to compute it. Besides, it fulfills the cocycle identity
\be
\omega_{h,k}\,\omega_{g,hk}=\omega_{gh,k}\,\omega_{g,h},
\ee
which ensures the associativity of the product of three operators,
$(T_{g}T_{h})T_{k}=T_{g}(T_{h}T_{k})$.
As operators acting on the particle's Hilbert space ${\cal H}$, they
generate an algebra which we denote by $\mathrm{C}_{\omega}[G]$, known
in mathematics as the twisted group algebra of $G$.

$T_{g}$ is not unique since one can always multiply $\phi_{g;\,i}$
by a constant $\alpha_{g}$. This changes the cocycle $\omega$ by a
coboundary,
\be
\omega_{g,h}\rightarrow\omega_{g,h}\frac{\alpha_{g}\alpha_{h}}{\alpha_{gh}}.
\label{alphapart}
\ee
Besides, if we change the bundle to an equivalent one, the
operators $T_{g}$ take the same form with modified $\phi_{g;\,i}$.
\medskip

In physics, the operators $T_{g}$ are expected to be symmetries of the
theory, which means that they commute with propagation. Though this
follows from the commutation of $T_{g}$ with $\nabla$, it is
instructive to derive this in the path integral framework. To this aim,
let us express the commutation relation $T_{g}K=KT_{g}$ using the
local forms of $T_{g}$ and $K$ given in \eqref{localK} and
\eqref{localT}. Starting with $\psi_{i}$, the local expression
for $KT_{g}\psi$ reads
\begin{equation}
\sum_{i}\int dx\,\rho_{i}(x)\, K_{ji}(y,x)\phi_{g;\,i}(x)\psi_{i}(x\!\cdot\!g),
\end{equation}
whereas $T_{g}K\psi$ yields
\begin{equation}
\phi_{g;\,j}(y)\sum_{i}\int dx\,\rho_{i}(x)\, K_{ji}(y\!\cdot\!g,x)
\psi_{i}(x)
=
\sum_{i}\int dx\,\rho_{i}(x)\,\phi_{g;\,j}(y) K_{ji}(y\!\cdot\!g,x\!\cdot\!g)
\psi_{i}(x\!\cdot\!g).
\end{equation}
To obtain this equality, we have performed a change of variable
$x\rightarrow x\!\cdot\!g$, assuming that the measure (usually associated
to the Riemannian metric involved in the kinetic term) and the
partition of unity are $G$-invariant.

Accordingly, the commutation of $T_{g}$ and $K$ follows if we have
\begin{equation}
K_{ji}(y\!\cdot\!g,x\!\cdot\!g)=
\phi_{g;\,i}^{-1}(y)K_{ji}(y,x)\phi_{g;\,i}(x).\label{propcom}
\end{equation}
Using \eqref{localpath}, the propagator $K_{ji}(y\!\cdot\!g,x\!\cdot\!g)$
is expressed as a path integral
\begin{equation}
K_{ji}(y\!\cdot\!g,x\!\cdot\!g)=\mathop{\int[D\varphi']}
\limits_{\begin{subarray}{c}
\varphi'(a)=x\cdot g\cr\varphi'(b)=y\cdot g
\end{subarray}}\,
\mathrm{e}^{-S[\varphi']}{\cal A}_{ji}[\varphi'].
\end{equation}
Then, we perform the change of variable $\varphi'=\varphi\!\cdot\!g$,
assuming that the measure and the kinetic terms $S[\varphi]$ of the
action are genuinely invariant under $G$,
\begin{equation}
K_{ji}(y\!\cdot\!g,x\!\cdot\!g)=\mathop{\int[D\varphi]}
\limits_{\begin{subarray}{c}
\varphi(a)=x\cr\varphi(b)=y
\end{subarray}}\,
\mathrm{e}^{-S[\varphi]}{\cal A}_{ji}[\varphi\!\cdot\!g].
\end{equation}
Then, \eqref{propcom} follows immediately from the transformation law
of the magnetic amplitude,
\be
\cA_{ji}[\varphi\!\cdot\!g]=
\phi_{g;\, j}^{-1}(y)\,\cA_{ji}[\varphi]\,\phi_{g;\,i}(x)\label{gApart}.
\ee
This last relation is easily checked using \eqref{phipart} and the
definition of the magnetic amplitude \eqref{holo}.

Alternatively, we could have determined the phase $\phi_{g;\,i}$ by
comparing the magnetic amplitude of $\varphi$ and
$\varphi\!\cdot\!g$. Then, we define $\phi_{g;\,i}$ such that
\eqref{gApart} holds. This provides the phase that must
accompany the pull-back action in the definition of $T_{g}$ in such a
way that it commutes with the propagator. This is the line of thought
we shall adopt for a string since it is versatile enough to also encompass
processes involving interactions.
\medskip

All the previous discussion refers to the quantum theory of a particle in
${\cal M}$ in the presence of a $G$-invariant magnetic field $B$. This
can serve as a basis for the construction of the analogous theory on the
quotient space ${\cal M}/G$, when we assume that the action is
free. Starting with the Hilbert space ${\cal
  H}$ we have constructed on ${\cal M}$, one would define the theory on
${\cal M}/G$ out of the wave functions $\psi\in{\cal H}$ that are
invariant under the action of $G$.  Because quantum mechanical states
only involve wave functions defined up to a phase, we only require
the invariance condition to hold up to a constant phase $\alpha_{g}\in U(1)$,
\begin{equation}
T_{g}\psi=\alpha_{g}\psi\quad\mathrm{for}\,\,\mathrm{any}\,\, g\in
G.\label{uni}
\end{equation}
This is incompatible with a non trivial projective action since a further
application of $T_{h}$ implies that $\omega$ is trivial,
\begin{equation}
\omega_{h,g}=\frac{\alpha_{h}\alpha_{g}}{\alpha_{gh}}.
\end{equation}
Therefore, the cohomology class of $\omega$ is an obstruction to the
construction of the theory on the quotient ${\cal M}/G$. This
difficulty can be bypassed if we define the physical states on the
quotient out of multidimensional subspaces in ${\cal H}$. Any state
would be an irreducible subrepresentation of the algebra generated by
the $T_{g}$'s instead of the unidimensional ones given in
\eqref{uni}. This point of view has been successfully applied to
abelian Maxwell-Chern-Simons theory for non integral values of the coupling
$k$ in \cite{polychronakos}. In this case, the role of the operators $T_{g}$
is played by the large gauge transformations of the space manifold.
\medskip

Abelian Maxwell-Chern-Simons theory is intimately tied up with the motion of a
particle on the plane in an external magnetic field, invariant under
some lattice translation. Therefore, let us consider a particle in
${\Bbb R}^{D}$ in an external magnetic field $B$, invariant under the group
$G={\Bbb Z}^{D}$ acting by translation. Because  $B=\d A$ is exact, the
bundle can be chosen to be trivial and the wave
functions are globally defined. In this case, $T_{g}$ takes the simple form
\be
T_{g}\psi(x)=
\mathrm{e}^{-\mathrm{i}\int_{x_{0}}^{x}\lp
  g^{\ast}A-A\rp}\,\psi(x\!\cdot\!g),
\label{trivial}
\ee
where we integrate along any path joining $x$ to a fixed reference
point $x_{0}$. Note that $g^{\ast}A-A$ is closed and
$\mathrm{H}^{1}({\cal M},U(1))$
trivial, so that the integral does not depend on the choice of the
path. Any change of the base point $x_{0}\rightarrow x'_{0}$
corresponds to a multiplication by
$\alpha_{g}=\mathrm{e}^{-\ii\int_{x_{0}}^{x'_{0}}g^{\ast}A-A}$.


Because $B$ is a $G$-invariant closed 2-form, it
can always be written using a Fourier series expansion as $B=B'+\d A'$,
with $B'$ constant and $A'$ invariant under lattice
translations. Therefore, we restrict our analysis to constant 2-forms
\be
B=\frac{1}{2}\,B_{\mu\nu}\,dx^{\mu}\wedge dx^{\nu},
\ee
for which $A$ can be chosen as
\be
A=\frac{1}{2}\, B_{\mu\nu}x^{\mu}\, dx^{\nu}.
\ee
In this gauge,
\be
T_{g}\psi(x)=\mathrm{e}^{-\frac{\ii}{2}B_{\mu\nu}g^{\mu}x^{\nu}}\,\psi(x\!\cdot\!g)
\label{torexplicit}
\ee
with $g\in\dZ^{D}$. These are nothing but the standard magnetic
translations, obeying the multiplication law
\begin{equation}
T_{g}T_{h}=\mathrm{e}^{\frac{\mathrm{i}}{2}B_{\mu\nu}g^{\mu}h^{\nu}}\;T_{gh}.
\end{equation}
The algebra they generate is nothing but a noncommutative torus.

This example also illustrates by explicit computations some of the
features of the general theory. The magnetic amplitude for the path is
\begin{equation}
\mathrm{e}^{\mathrm{i}\int\varphi^{\ast}A}=
\mathrm{e}^{\frac{\mathrm{i}}{2}\int dt\, B_{\mu\nu}\frac{d\varphi^{\mu}}{dt}\varphi^{\nu}}
\end{equation}
and fulfills
\begin{equation}
\mathrm{e}^{\mathrm{i}\int(\varphi\cdot g)^{\ast}A}=
\mathrm{e}^{\frac{\ii}{2}B_{\mu\nu}g^{\mu}y^{\nu}}\,
\mathrm{e}^{\mathrm{\ii}\int\varphi^{\ast}A}\,
\mathrm{e}^{-\frac{\ii}{2}B_{\mu\nu}g^{\mu}x^{\nu}},
\end{equation}
in accordance with \eqref{gApart}. The euclidian kinetic term
$\int d\tau\, \frac{m}{2}\lp\frac{dx}{d\tau}\rp^{2}$ as well
as the path integral measure are invariant. In two dimensions, the
resulting gau\ss ian path integral can be evaluated explicitly
\begin{equation}
\mathop{\int[\mathrm{D}\varphi]}
\limits_{\begin{subarray}{c}
\varphi(b)=y\cr\varphi(a)=x
\end{subarray}}
\,\mathrm{e}^{-\int \frac{m}{2}
\left(\frac{\mathrm{d}\varphi}{\mathrm{d}t}\right)^{2}+
\mathrm{i}\int\varphi^{\ast}A}=
\frac{B}{4\pi\sinh\left(\frac{B\Delta\tau}{m}\right)}
\exp\left\{-\frac{B}{4}\coth\left(\frac{B\Delta\tau}{m}\right)
\left(x-y\right)^{2}+\ii\frac{B}{2}x\wedge y
\right\}
\end{equation}
with $x\wedge y=x_{1}y_{2}-x_{2}y_{1}$ and $\Delta\tau=a-b$ being the
euclidian time interval. On this simple example, one can check that
the propagator fulfills \eqref{propcom} under the lattice translation given by
\eqref{torexplicit}. Though this example involves an infinite group so
that it does not fit in our general theory, it is the simplest example
available with globally defined fields. Indeed, if we assume that $G$
is finite and the fields globally defined, then averaging over $G$
renders any potential $G$-invariant so that the phase $\phi_{g}$
disappears.

\section{Higher gauge fields and their invariance}

\subsection{The tricomplex}

As we have seen for a particle moving in an external magnetic field
invariant under a finite group $G$, the analysis of the invariance of the
gauge potentials and the wave functions interplays the de Rham
differential with \Cech{} and group coboundaries. In the case of a
string, the corresponding magnetic field is a 3-form $H$ with locally
defined 2-form potentials $B_{i}$ as well as 1-forms $B_{ij}$ and transition
functions $f_{ijk}$. In order to handle this multiplet of potentials and
its interplay with the group action, it is convenient to combine the de Rham,
\Cech{} and group cohomology into a single object which we now define.

To begin with, assume that  $\cM$ has been covered with a good
invariant cover $\la U_{i}\ra$. We define a tricomplex whose cochains
$C_{p,q,r}$ are de Rham forms of degree $p$, defined on $(q+1)$-fold
intersections, $U_{i_{0}}\cap\dots\cap U_{i_{q}}$ and functions of $r$
group indices. These forms are real valued for $p>0$ and $U(1)$-valued
for $p=0$. In order not to single out the case $p=0$ in all the
forthcoming formulae, we shall always use
the additive notation when $p$ is generic. It is
self-understood that for $p=0$, addition and opposites have to be traded for
multiplication and inverses. Besides, elements of $C_{p,q,r}$ are
completely antisymmetric in the indices labelling the open sets involved
in the intersection, where for $p=0$ the antisymmetry involves the
inverse instead of the opposite.

The differential $\mathrm{d}$ in the $p$-direction is the de Rham one
for $p>0$ and
$\ii\,\d\log$ for $p=0$. In the $q$-direction, we take the \u{C}ech
coboundary defined as
\be
(\check\delta c)_{i_{0}\dots i_{q}}=
\sum_{k=0}^{q}(-1)^{k}c_{i_{0}\dots,\check{i}_{k}\dots,i_{q}}
\ee
where $\check{i}_{k}$ means that the index $i_{k}$ has been
omitted. For low values of $q$, we have
\begin{equation}
\left\{
\begin{array}{rclcl}
(\check\delta c)_{ij}&=&c_{j}-c_{i}&\mathrm{on}&U_{i}\cap U_{j},\\
(\check\delta c)_{ijk}&=&c_{jk}-c_{ik}+c_{ij}
&\mathrm{on}&U_{i}\cap U_{j}\cap U_{k}\\
(\check\delta c)_{ijkl}&=&c_{jkl}-c_{ikl}+c_{ijl}-c_{ijk}
&\mathrm{on}&U_{i}\cap U_{j}\cap U_{k}\cap U_{l}.
\end{array}
\right.
\end{equation}
The first equation is a measure of how far is $c_{i}$ from a global
object whereas the other ones compare objects defined on multiple
intersections.

Finally, in the $r$-direction, we take the group coboundary
operator
\be
(\delta c)_{g_{0},\dots,g_{r}}=g_{0}^{\ast}c_{g_{1},\dots, g_{r}}
+\sum_{k=1}^{r}(-1)^{k}\,c_{g_{0},\dots,g_{k-1}g_{k},\dots,g_{r}}
+(-1)^{r+1}c_{g_{0},\dots, g_{r-1}},
\ee
where $g^{\ast}$ is the pullback action on forms. For low values of $r$,
one has
\begin{eqnarray}
(\delta c)_{g}&=&g^{\ast}c-c,\\
(\delta c)_{g,h}&=&g^{\ast}c_{h}-c_{gh}+c_{g},\label{group2}\\
(\delta c)_{g,h,k}&=&g^{\ast}c_{h,k}-c_{gh,k}+c_{g,hk}-c_{g,h}.\label{group3}
\end{eqnarray}
The first equation quantifies the lack of invariance of $c$ under the
action of $g$. The other equations  appear in physics in
the definition of representations up to a phase \cite{jackiw}. Indeed,
if we represent the group $G$ on some functions by
$T_{g}\psi(x)=\rho_{g}(x)\psi(x\!\cdot g)$, the multiplication law
of $G$ is fulfilled iff $\rho$ satisfies \eqref{group2}. The third
equation enters into the definition of projective representations:
$T_{g}T_{h}=\omega_{g,h}T_{gh}$ defines an associative multiplication
law iff $\omega$ obeys \eqref{group3}. Because most of the
multiplication laws encountered in physics involve the operator
multiplication, a failure of associativity seldom occurs. Let us also
note that we have defined the group coboundary $\delta$ for
differential forms on which $G$ acts by pull-back but the same definition is
valid for an arbitrary representation of $G$.
Obviously, the three differentials commute so that they define a tricomplex.

\begin{figure}[h]
\[ \xymatrix @R=1.0pc @C=2.0pc {
            & \bullet \ar[rr] & & \bullet \\
            \bullet \ar[rr] \ar[ru] & & \bullet \ar[ru] & \\
            & \bullet \ar'[r][rr] \ar'[u][uu] & & \bullet\ar[uu] \\
            \bullet \ar[rr]_{\mathrm{d}/p} \ar[uu]^{\delta/r} \ar[ru]^{\check\delta/q} & &\bullet \ar[uu] \ar[ru] &  }
\]
\caption{Directions of the coboundaries.}
\end{figure}
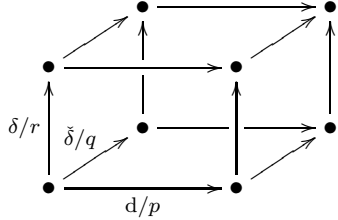

Thus, for any fixed value of $r$, we have a \Cech-de Rham bicomplex, out of
which we form a total complex
\be
C_{r,s}^{\mathrm{tot}}=\bigoplus_{p+q=s}C_{p,q,r},
\ee
equipped with the Deligne differential defined by
\begin{equation}
\lp\cD c\rp_{p,q,r}=
(-1)^{q}\mathrm{d}c_{p-1,q,r}+(-1)^{q-1}\check{\delta}c_{p,q-1,r}
\end{equation}
for $c=(c_{p,q,r})_{r,\,p+q=s}$ in $C_{r,s}^{\mathrm{tot}}$. We define
$\mathrm{d}c_{p-1,q,r}$ (resp. $\check{\delta}c_{p,q-1,r}$) as 0 when
$p=0$ (resp. $q=0$). Obviously, $\cD$ squares to 0 and commutes with
the group coboundary $\delta$. Up to multiplication by a global sign on
$C_{r,s}^{\mathrm{tot}}$, this \Cech-de Rham
bicomplex is identical to the one presented
in \cite{gawedzki}, so that it follows that they share the same
cohomology.
\medskip

Using the tricomplex, the analysis we have performed for a particle takes
a simple form. Define
\be
{\bf B}=\lp B_{i},0,1\rp\in C_{0,2}^{\mathrm{tot}}.
\ee
Because the $B$-field is closed and globally defined, one has $\cD
{\bf B}=0 $. The relations defining the gauge fields and the
transition functions read $\cD {\bf A}={\bf B}$, with
\be
{\bf A}=\lp A_{i},f_{ij}\rp\in C_{0,1}^{\mathrm{tot}}.
\ee
Gauge transformations are given by $\eta=(\eta_{i})\in
C_{0,0}^{\mathrm{tot}}$ and act on ${\bf A}$ as ${\bf A}\rightarrow
{\bf A}+\cD \eta$. This summarizes the definition of a line bundle
with connection as well as their gauge equivalence.

Now assume that the group $G$ preserves the  $B$-field and that the degree one
cohomology of the total \Cech-de Rham complex is trivial, as is the
case if $\mathrm{H}^{1}(\cM,U(1))$ is trivial. Because
${\bf B}$ is invariant, $\delta {\bf B}=0$, but ${\bf A}$ needs not to
be invariant. However,
\begin{equation}
\cD\delta {\bf A}=\delta \cD {\bf A}=\delta {\bf B}=0,\label{deltaA}
\end{equation}
so that the triviality of the degree 1 cohomology implies the
existence of $\Phi\in C_{1,0}^{\mathrm{tot}}$ such that
$\delta {\bf A}=\cD\Phi$. Evaluating both sides
of \eqref{deltaA} on $g\in G$ yields $g^{\ast}{\bf A}={\bf
  A}+\cD\Phi_{g}$. With $\Phi_{g}=\lp \phi_{g;\,i}\rp$, it reproduces
\eqref{phipart}. Finally, we define $\omega=\delta\Phi$ that fulfills
\be
\cD\omega=\cD\delta \Phi=\delta\cD\Phi=\delta^{2}{\bf A}=0.
\ee
It shows that $\omega$ is constant ($\mathrm{d}\omega=0$) and globally
defined ($\check{\delta}\omega=0$). By definition, $\omega$ also
satisfies $\delta\omega=0$ so that it is a group 2-cocycle. If the
degree one cohomology of the \Cech-de Rham bicomplex is not trivial,
then there is an obstruction to solving $\cD\delta{\bf A}=0$, which
corresponds to the fact that a line bundle and its pull-back by $g$
are not necessary isomorphic.

The ambiguity in the definition of $\Phi$ is parametrized by
$\alpha\in C^{\mathrm{tot}}_{1,0}$
such that $\cD\alpha=0$ which means that $\alpha$ is a globally
defined constant group cochain. This induces the changes
\begin{equation}
\left\{
\begin{array}{rcl}
\Phi&\rightarrow&\Phi\alpha\cr
\omega&\rightarrow&\omega\delta\alpha.
\end{array}
\right.
\label{ambiguityalpha}
\end{equation}
Besides, the magnetic amplitude defined in (\ref{holo}) for a path
$\varphi$ joining $x$ to $y$ can be written symbolically as
\be
\cA_{ji}[\varphi]=\lb \mathrm{e}^{\ii\int_{x}^{y}{\bf A}}\rb_{ji}.
\ee
The use of a triangulation is self-understood and it only depends on the
the open sets $U_{i}$ and $U_{j}$ used to cover the endpoints, as
emphasized by the notation $[\dots]_{ji}$. Gauge
transformations act as
\be
\lb \mathrm{e}^{\ii\int_{x}^{y}{\bf A}}\rb_{ji}\rightarrow
\lb \mathrm{e}^{\ii\int_{x}^{y}{\bf A}+\cD\eta}\rb_{ji}=
\left[{\eta}^{-1}(y)\right]_{j}\,\lb \mathrm{e}^{\ii\int_{x}^{y}{\bf A}}\rb_{ji}
\,\left[{\eta}(x)\right]_{i},
\ee
with $\left[\eta(x)\right]_{i}=\eta_{g;\,i}(x)$.  The action of $g\in
G$ on the amplitude reads
\be
\cA_{ji}[\varphi\!\cdot\!g]=
\lb \mathrm{e}^{\ii\int_{x}^{y}{g^{\ast}\bf A}}\rb_{ji}=
\left[\Phi_{g}^{-1}(y)\right]_{j}
\lb \mathrm{e}^{\ii\int_{x}^{y}{\bf A}}\rb_{ji}
\left[\Phi_{g}(x)\right]_i,
\ee
with $\left[\Phi_{g}(x)\right]_{i}=\phi_{g;\,i}(x)$. This notation
encodes the formal analogy between cochains of the \Cech-de Rham
bicomplex and ordinary differential forms.

\subsection{2-form gauge potentials}

Consider now a closed 3-form $H$ such that its de Rham cohomology
class belongs to $\mathrm{H}^{3}(\cM,2\pi\dZ)$, which means that its integral
over any closed 3-manifold in $\cM$ belongs to $2\pi\dZ$.
Under these assumptions, one can find locally defined forms of lower
degrees such that
\be
\la
\begin{array}{lcrcl}
H_{i}&=&\d B_{i}\quad&\mbox{on}&\quad U_{i},\cr
B_{j}-B_{i}&=&\d B_{ij}\quad&\mbox{on}&\quad U_{i}\cap U_{j},\cr
B_{jk}-B_{ik}+B_{ij}&=&\ii\,\d\log f_{ijk}
\quad&\mbox{on}&\quad U_{i}\cap U_{j}\cap U_{k},\cr
f_{jkl}(f_{ikl})^{-1}f_{ijl}(f_{ijk})^{-1}&=&1
\quad&\mbox{on}&\quad U_{i}\cap U_{j}\cap U_{k}\cap U_{l},\label{gerb}
\end{array}
\right. \ee where all forms are antisymmetric in their indices. On
the mathematical side, these relations define a {\it gerb with
connection} \cite{hitchin}, but we will simply refer to these fields
as 'the $B$-fields'. The last equation requires the de Rham
cohomology class of $H$ to be integral. The prototypical example is
the magnetic part of the action of the Wess-Zumino-Witten model that
describes strings moving on the group $\mbox{SU(N)}$ in the presence
of the 3-form $H=\frac{k}{12\pi}\mbox{Tr}(g^{-1}dg)$
\cite{gawedzki}.


Given $H$, the $B$-fields are not unique. One can gauge transform them as
\be
\la
\begin{array}{lcl}
B_{i}&\rightarrow&B_{i}+\d\Lambda_{i},\cr
B_{ij}&\rightarrow&B_{ij}+\Lambda_{j}-\Lambda_{i}-\ii\,\d\log\eta_{ij},\cr
f_{ijk}&\rightarrow&f_{ijk}\,\eta_{jk}^{-1}\,\eta_{ik}\,\eta_{ij}^{-1},
\end{array}
\right.
\label{gaugestring1}
\ee
while preserving \eqref{gerb}. Note that the gauge transformation now
involves real 1-forms $\Lambda_{i}$ and $U(1)$ valued functions
$\eta_{ij}$. Besides, there is a gauge transformation of the gauge
transformation since
\be
\la
\begin{array}{lcl}
\Lambda_{i}&\rightarrow&\Lambda_{i}+\ii\,\d\log\xi_{i} ,\cr
\eta_{ij}&\rightarrow&\eta_{ij}\,\xi_{j}\,(\xi_{i})^{-1}
\end{array}
\right.
\ee
has no effect on the $B$-fields. For the sake of brevity, we call the
latter {\it residual gauge transformations}.

To write these equations using the \Cech-de Rham bicomplex, let us define
\be
{\bf H}=\lp H_{i},0,0,1\rp\in C_{0,3}^{\mathrm{tot}},
\ee
which fullfils $\cD {\bf H}=0$ because $H$ is closed and globally
defined. The system of equations \eqref{gerb} defining the $B$-fields
are conveniently written as ${\bf H}=\cD {\bf B}$, with
\be
{\bf B}=\lp B_{i},B_{ij},f_{ijk}\rp\in C_{0,2}^{\mathrm{tot}}.
\ee
Gauge transformations, gathered into $\Lambda=\lp
\Lambda_{i},\eta_{ij}\rp\in C_{0,1}^{\mathrm{tot}}$, act as
${\bf B}\rightarrow {\bf B}+ \cD\Lambda$, whereas residual
gauge transformations $\xi=(\xi_{i})\in C_{0,0}^{\mathrm{tot}}$ act on the
gauge transformations as $\Lambda\rightarrow\Lambda+\cD\xi$.
\medskip

Consider now a finite group $G$ acting on $\cM$ and leaving the $H$-field
invariant. We assume there is no cohomology
in the total \Cech -de Rham bicomplex in degree 1 and 2, as is the
case if $\mathrm{H}^{1}(\cM,U(1))$ and
$\mathrm{H}^{2}(\cM,U(1))$ are trivial. This holds for simply connected simple Lie groups like $\mbox{SU}(N)$ that
are target spaces of the Wess-Zumino-Witten models. We repeat the
same pattern as for a particle, with one extra step.

The invariance of ${\bf H}$ translates into $\delta {\bf H}=0$, which
implies
\be
\cD \delta {\bf B}=\delta \cD {\bf B}=\delta {\bf H}=0.
\ee
Because the degree 2 cohomology of the \Cech{}-de Rham bicomplex is
trivial, there exists an  ${\bf A}\in C_{1,1}^{\mathrm{tot}}$ such that
\be
\delta {\bf B}=\cD {\bf A}.
\ee
${\bf A}$ is defined up to $\cD\Theta$ with $\Theta\in
C_{1,0}^{\mathrm{tot}}$. By the same token,
\be
\cD\delta {\bf A}=\delta\cD {\bf A}=\delta^{2}{\bf B}=0,
\ee
so that one can find $\Phi\in C_{2,0}^{\mathrm{tot}}$ such
that $\delta {\bf A}=\cD \Phi$.
$\Phi$ is defined up to an element
$\alpha\in C_{2,0}^{\mathrm{tot}}$ such that $\cD\alpha=0$,
which is a globally defined constant group 2-cochain. Finally,
$\omega=\delta \Phi$ is a constant and globally defined element of
$C_{3,0}^{\mathrm{tot}}$ since
\begin{equation}
\cD\omega=\cD\delta\Phi=\delta\cD\Phi=\delta^{2}{\bf A}=0.
\end{equation}
Because of
\begin{equation}
\delta\omega=\delta^{2}\Phi=0,
\end{equation}
it is a group 3-cocycle. Despite its form, it is important to note
that $\omega$ is not a coboundary. Indeed, it would be so only if
$\Phi$ were a constant, which is usually not the case. The same remark
applies to the 2-cocycle pertaining to the particle in a magnetic
field, see \eqref{2cocycle}.

\begin{figure}[h]\centering
\includegraphics[width=9cm]{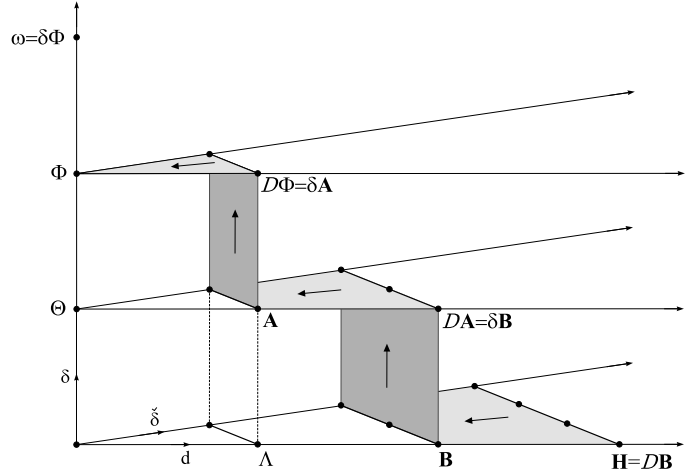}
\caption{Derivation of ${\bf B}$, ${\bf A}$, $\Phi$ and $\omega$.}
\label{_tricomplex}
\end{figure}

The definitions of ${\bf B}$, ${\bf A}$, $\Phi$ and $\omega$
can be recovered from figure \ref{_tricomplex}.
%
We always move towards the $r$ axis by inverting $\cD$ up to gauge
transformations, and to the top using $\delta$.

It is useful to evaluate the group cochains explicitly, so that all our
relations read
\be
\la
\begin{array}{rcl}
\cD {\bf A}_{g}&=&g^{\ast}{\bf B}-{\bf B},\cr
\cD \Phi_{g,h}&=&g^{\ast}{\bf A}_{h}-{\bf A}_{gh}+{\bf A}_{g},\cr
\omega_{g,h,k}&=&g^{\ast}\Phi_{h,k}\,\Phi^{-1}_{gh,k}\,\Phi_{g,hk}\,\Phi^{-1}_{g,h}.
\end{array}
\right.
\label{groupexplicit}
\ee
The first equation means that ${\bf A}_{g}$ is a gauge transformation from
${\bf B}$ to $g^{\ast}{\bf B}$. Then, $(gh)^{\ast}{\bf B}$ can be obtained from ${\bf B}$
either by ${\bf A}_{gh}$, or by ${\bf A}_{g}+g^{\ast}{\bf A}_{h}$. Therefore, the two
gauge transformations must be related by a residual gauge transformation
$\Phi_{g,h}$. This is encoded in the second
equation. The last equation arises from the two ways of combining two
residual gauge transformations from ${\bf A}_{ghk}$ to
${\bf A}_{g}+h^{\ast}{\bf A}_{g}+(gh)^{\ast}{\bf A}_{k}$. Indeed, one can use either
\begin{equation}
{\bf A}_{ghk}\xrightarrow{\Phi_{g,hk}}{\bf A}_{g}+g^{\ast}{\bf A}_{hk}
\xrightarrow{g^{\ast}\Phi_{h,k}}{\bf A}_{g}+g^{\ast}{\bf A}_{h}+(gh)^{\ast}{\bf
  A}_{k}
\label{gauge1},
\end{equation}
or
\begin{equation}
{\bf A}_{ghk}\xrightarrow{\Phi_{gh,k}}{\bf A}_{gh}+(gh)^{\ast}{\bf A}_{k}
\xrightarrow{\Phi_{g,h}}{\bf A}_{g}+g^{\ast}{\bf A}_{h}+(gh)^{\ast}{\bf A}_{k}.
\label{gauge2}
\end{equation}
Since the two combinations of residual gauge transformations have the
same action on ${\bf A}_{ghk}$,
they differ by a constant which is $\omega_{g,h,k}$.

\begin{figure}[h]
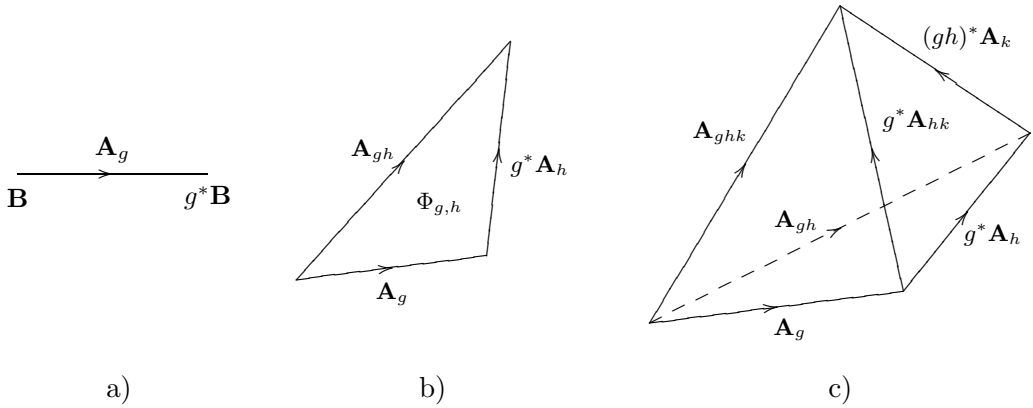
\centering
\begin{tabular}{ccc}
\parbox{3.5cm}{
\[ \xy 0;/r.15pc/:
(0,5)*{}="lu"; (40,5)*{}="ru"; "lu";"ru" **\dir{-}?(.5)*\dir{>};
(20,10)*{\mbox{\small ${\bf A}_g$}}; (0,0)*{\mbox{\small ${\bf
B}$}}; (40,0)*{\mbox{\small $g^\ast {\bf B}$}};
\endxy \]}
&\parbox{4cm}{
\[ \xy 0;/r.15pc/:
(0,0)*{}="lu";
(40,5)*{}="ru";
(45,50)*{}="o";
"lu";"ru" **\dir{-}?(.5)*\dir{>};
"ru";"o" **\dir{-}?(.5)*\dir{>};
"lu";"o" **\dir{-}?(.5)*\dir{>};
(20,-3)*{\mbox{\footnotesize ${\bf A}_g$}};
(51,24)*{\mbox{\footnotesize $g^\ast {\bf A}_h$}};
(16,27)*{\mbox{\footnotesize ${\bf A}_{gh}$}};
(30,16)*{\mbox{\footnotesize $\Phi_{g,h}$}};
\endxy \]}
&\parbox{6cm}{
\[ \xy 0;/r.2pc/:
(0,0)*{}="lu";
(40,5)*{}="ru";
(60,30)*{}="rm";
(30,50)*{}="o";
"lu";"ru" **\dir{-}?(.5)*\dir{>};
"ru";"rm" **\dir{-}?(.5)*\dir{>};
"rm";"o" **\dir{-}?(.5)*\dir{>};
"lu";"o" **\dir{-}?(.5)*\dir{>};
"lu";"rm" **\dir{--}?(.5)*\dir{>};
"ru";"o" **\dir{-}?(.5)*\dir{>};
(22,-1)*{\mbox{\footnotesize ${\bf A}_g$}};
(54,14)*{\mbox{\footnotesize $g^\ast {\bf A}_h$}};
(50,45)*{\mbox{\footnotesize $(gh)^\ast {\bf A}_k$}};
(11,30)*{\mbox{\footnotesize ${\bf A}_{ghk}$}};
(23,16)*{\mbox{\footnotesize ${\bf A}_{gh}$}};
(42,32)*{\mbox{\footnotesize $g^\ast {\bf A}_{hk}$}};
\endxy \]}\\ a)&b)&c)
\end{tabular}
\caption{Geometric illustration for group cochains.}
\label{resi3}
\end{figure}

In figure \ref{resi3}, vertices correspond to $B$-fields, arrows to gauge
transformations ${\bf A}_{g}$, faces to residual gauge transformations
$\Phi_{g,h}$ and $\omega_{g,h,k}$ to the tetrahedron obtained by
gluing the two couples of triangles corresponding to \eqref{gauge1} and
\eqref{gauge2}.

Combining the gauge ambiguity on ${\bf B}$ as well as the ambiguities
in the definition of ${\bf A}$ and $\Phi$, we obtain the following transformations
\be
\la
\begin{array}{lcl}
{\bf B}&\rightarrow&{\bf B}+\cD\Lambda\cr
{\bf A}&\rightarrow&{\bf A}+\delta\Lambda+\cD\Theta,\cr
\Phi&\rightarrow&\Phi\,\delta\Theta\,\alpha,\cr
\omega&\rightarrow&\omega\,\delta\alpha.
\end{array}
\right.
\ee
It is worthwhile to notice that $\omega$ is unaffected by the gauge
transformations $\Lambda$ and $\Theta$, and is
only modified through a group coboundary by $\alpha$. Upon evaluating
the group indices, we get
\be
\la
\begin{array}{lcl}
{\bf B}&\rightarrow&{\bf B}+\cD\Lambda\cr
{\bf A}_{g}&\rightarrow&{\bf A}_g+g^{\ast}\Lambda-\Lambda+\cD\Theta_{g},\cr
\Phi_{g,h}&\rightarrow&
\Phi_{g,h}\;g^{\ast}\Theta_{h}\lp\Theta_{gh}\rp^{-1}\Theta_{g}\;\alpha_{g,h},\cr
\omega_{g,h,k}&\rightarrow&\omega_{g,h,k}\;
\alpha_{h,k}\lp\alpha_{gh,k}\rp^{-1}\alpha_{g,hk}\lp\alpha_{g,h}\rp^{-1}.
\end{array}
\right.
\label{gaugestring3}
\ee
As a convention, we always assume that all the group cochains are
normalized, i.e. they vanish whenever one of their arguments equals the
identity of the group. This is always possible by exploiting the gauge freedom.

Besides, it is helpful to display {\Cech} indices explicitly. We write
${\bf A}_{g}=\lp A_{g;\,i},f_{g;\,ij}\rp$ and
$\Phi_{g,h}=\lp\phi_{g,h;\,i}\rp$, so that the system of equations \eqref{groupexplicit} reads
\be
\la
\begin{array}{rcl}
\d A_{g;\,i}&=&g^{\ast}B_{i}-B_{i},\cr
A_{g;\,j}-A_{g;\,i}-\ii\,\d\log f_{g;\,ij}&=&g^{\ast}B_{ij}-B_{ij},\cr
(f_{g;\,jk})^{-1}f_{g;\,ik}(f_{g;\,ij})^{-1}&=&g^\ast f_{ijk}(f_{ijk})^{-1}\cr
\ii\,\d\log\phi_{g,h;\,i}&=&g^{\ast}A_{h;\,i}-A_{gh;\,i}+A_{g;\,i},\cr
\phi_{g,h;\,j}(\phi_{g,h;\,i})^{-1}&=&g^{\ast}f_{h;\,ij}(f_{gh;\,ij})^{-1}f_{g;\,ij},\cr
\omega_{g,h,k}&=&
g^{\ast}\phi_{h,k;\,i}\lp\phi_{gh,k;\,i}\rp^{-1}\phi_{g,hk;\,i}\lp\phi_{g,h;\,i}\rp^{-1}.
\end{array}
\right.
\label{cech}
\ee
Up to a different sign convention, these are the equations derived by
E. Sharpe in his analysis of discrete torsion \cite{sharpe}, at the
notable exception of the 3-cocycle $\omega$. As follows from our
discussion, the triviality of $\omega$ cannot be
obtained solely on the grounds of the invariance of $H$. However, it
is an essential consistency condition in order to build an
orbifold theory. We shall come back to this point at the end of
section \ref{orbifold}.

In the next section we will also have to use the equations obtained by
displaying {\Cech} indices for the gauge transformations,
\be
\la
\begin{array}{rclcl}
A_{g;\,i}&\rightarrow&A_{g;\,i}+g^{\ast}\Lambda_{i}-\Lambda_{i}
+\ii\,\d\log\theta_{g;\,i},\cr
f_{g;\,ij}&\rightarrow&f_{g;\,ij}\;
\theta_{g,j}(\theta_{g;\,i})^{-1}g^{\ast}\eta_{ij}(\eta_{ij})^{-1},\cr
\phi_{g,h;\,i}&\rightarrow&
\phi_{g,h;\,i}\;g^{\ast}\theta_{h;\,i}(\theta_{gh;\,i})^{-1}\theta_{g;\,i},
\end{array}
\right.
\label{gaugestring4}
\ee
with $\Theta=\lp \theta_{g;\,i}\rp$. The transformation law of
the components of ${\bf B}$ are identical to (\ref{gaugestring1}). 

\medskip

It is worthwhile to work out a simple example with globally
defined fields. This requires the use of an infinite
group and the simplest example is provided by the constant 3-form
\be
H=\frac{1}{6}\, H_{\mu\nu\lambda}\,dx^{\mu}\wedge dx^{\nu}\wedge dx^{\lambda}
\ee
on $\dR^{N}$ with the group $G=\dZ^{N}$. $B$, $A$ and $\phi$ are
globally defined differential forms which are given, in suitable gauges,
by
\be
\la
\begin{array}{rcl}
B&=&\scriptsize{\frac{1}{6}}H_{\mu\nu\lambda}x^{\mu}\,dx^{\nu}\wedge dx^{\lambda}\cr
A_{g}&=&\frac{1}{6}H_{\mu\nu\lambda}g^{\mu}x^{\nu}\, dx^{\lambda}\cr
\phi_{g,h}&=&
\exp
\ii\la\frac{1}{6}H_{\mu\nu\lambda}g^{\mu}h^{\nu}x^{\lambda}\ra\cr
\omega_{g,h,k}&=&
\exp
\ii\la\frac{1}{6}H_{\mu\nu\lambda}g^{\mu}h^{\nu}k^{\lambda}\ra
\label{globalfields}
\end{array}
\right..
\ee

\subsection{An interlude on the M-theory 3-form}

We have worked out the explicit construction for a particle and a
string, involving field strengths that are 2-forms and 3-forms. For
higher dimensional extended objects, the same method applies. Starting
with an $n$-form field strength which is invariant under
the group $G$, we derive using the tricomplex a sequence of
$k$-forms carrying $n-k-1$ group indices for $k=0,\dots,n-1$ and a
constant group $n$-cocycle. This should be useful in the
analysis of a supersymmetric theory like type IIA and IIB, whose
spectrum involves forms of various degrees coupling to the
branes. Analogously, this also applies to the M-theory 3-form $C$ and
readily  yields equations identical to those obtained in
\cite{sharpeMtheory} and \cite{seki} in case of a trivial $4$-cocycle,
as we now illustrate.

Let us start with an invariant globally defined closed 4-form $G$ that
stands for the field strength. From $G$, we build
\begin{equation}
{\bf G}=\left(G_{i},0,0,0,1\right)\in C_{0,4}^{\mathrm{tot}}.
\end{equation}
It admits a set of locally defined potentials that we gather into
\be
{\bf C}=\lp C_{i},C_{ij},C_{ijk},f_{ijkl}\rp\in C_{0,3}^{\mathrm{tot}},
\ee
standing for the locally defined 3-form potentials $C_{i}$ and lower
degree forms needed to glue them on non
trivial intersections. The relation between the potentials and the
field strength is summarized by
\begin{equation}
\cD{\bf C}={\bf G}.\label{_2-gerb}
\end{equation}
There is a 2-form gauge invariance with gauge transformations $\Lambda\in
C_{0,2}^{\mathrm{tot}}$,
\begin{equation}
{\bf C}\rightarrow{\bf C}+\cD\Lambda
\end{equation}
as well as two levels of residual gauge transformations $\Xi\in
C_{0,1}^{\mathrm{tot}}$ and $\chi\in C_{0,0}^{\mathrm{tot}}$ acting as
\begin{equation}
\Lambda\rightarrow\Lambda+\cD\Xi\quad\mathrm{and}\quad
\Xi\rightarrow\Xi+\cD\chi.
\end{equation}
In the mathematical language, the five equations obtained by expanding
\eqref{_2-gerb} define a {\it 2-gerb with connection}. Such an object is
fundamental in M-theory since it couples to the worldvolume of a
membrane in the same way that a $B$-field couples to a string.

Let us now assume that the field strength ${\bf G}$ is invariant under
a group $G$ so that $\delta {\bf G}=0$. We further assume that $\cM$
is such that the degree $k$ cohomology groups of the \Cech-de Rham
bicomplex are trivial for $k=1,2,3$. Thus, one can define ${\bf B}\in C_{1,2}^{\mathrm{tot}}$, ${\bf A}\in
C_{2,1}^{\mathrm{tot}}$ and
$\Phi\in C_{3,0}^{\mathrm{tot}}$ such that $\cD {\bf B}=\delta
{\bf C}$, $\cD {\bf A}=\delta {\bf B}$ and $\cD\Phi=\delta {\bf A}$ as well as
$\omega=\delta\Phi$ which is a globally defined constant
($\cD\omega$=0) group 4-cocycle ($\delta\omega$=0). The rationale of
this derivation is to move from ${\bf G}$ to the 4-cocycle $\omega$ in
the tricomplex by inverting ${\cal D}$ up to gauge transformations to
decrease the total \Cech-de Rham degree and applying $\delta$ to get
higher group cochains.

Upon displaying explicitly the group indices, these equations read
\be
\la
\begin{array}{rcl}
\cD {\bf B}_{g}&=&g^{\ast}{\bf C}-{\bf C},\cr
\cD {\bf A}_{g,h}&=&g^{\ast}{\bf B}_{h}-{\bf B}_{gh}+{\bf B}_{g},\cr
\cD\Phi_{g,h,k}&=&
g^{\ast}{\bf A}_{h,k} - {\bf A}_{gh,k} + {\bf A}_{g,hk} - {\bf A}_{g,h},\cr
\omega_{g,h,k,l}&=&
g^{\ast}\Phi_{h,k,l}\lp\Phi_{gh,k,l}\rp^{-1}\Phi_{g,hk,l}\lp\Phi_{g,h,kl}\rp^{-1}
\Phi_{g,h,k}.
\end{array}
\right.
\ee
If we assume that $\omega$ is trivial and display all \Cech{} indices,
then these equations are identical to the ones in
\cite{sharpeMtheory} and \cite{seki}. Besides, there are various gauge ambiguities in
the definitions of the fields ${\bf B}$, ${\bf A}$ and $\Phi$ as well as an extra
constant phase in the definition of $\Phi$ alone. This is a group
3-cochain that changes $\omega$ by a coboundary. If we require that
$\omega$ remains unchanged, it is a 3-cocycle which is identified in
\cite{sharpeMtheory} with the M-theory analogue of discrete torsion.

\section{Propagating string}

\subsection{Magnetic amplitudes and string wave functions}

As for a particle, a closed string propagating on ${\cal M}$ in a background
3-form $H$ can be described in the functional integral approach by a magnetic
amplitude. More precisely, the string propagator $K$ can be derived in
perturbation theory from CFT as\footnote{To be more precise one has to
further integrate over the moduli and sum over the genera of ${\cal S}$}
\begin{equation}
K(Y,X)
=\mathop{\int[D\varphi]}\limits_{\varphi(\partial{\cal S})=X^{\ast}\cup Y}
\,e^{-S[\varphi]}{\cal A}[\varphi],\label{stringpath}
\end{equation}
where $\varphi$ is a field from a two dimensional surface ${\cal S}$
into ${\cal M}$ whose boundary values coincide with the string initial
and final positions $X$ and $Y$. Note that $X$ and $Y$ may both have
several connected components and the topology of ${\cal S}$ encodes
all the possible interactions. Besides,  $K(Y,X)$ fulfills a set of
axioms \cite{segal} and provides the evolution of the string wave
function according to
\begin{equation}
\Psi(X)\rightarrow \Psi'(Y)=\int [DX]\,K(Y,X)\Psi(X),
\end{equation}
where $\Psi$ and $\Psi'$ are functionals of the string's embedding
in space-time. In the following, we confine ourselves to the
simplest topologies for ${\cal S}$:  a cylinder (free string
propagation) and a pair of pants (tree level decay of a string).

The magnetic amplitude ${\cal A}[\varphi]$ encodes all the couplings
to the potentials of the external $H$-field and defines, together
with $S[\varphi]$ a conformally invariant field theory on ${\cal
S}$. Recall that the typical example we have in mind is the WZW
model with ${\cal M}=\mathrm{SU(N)}$ and
$H=\frac{k}{12\pi}\mathrm{Tr}\left(
  gdg^{-1}\right)^{3}$ the Maurer-Cartan 3-form.
In this case, $H$ is closed but not exact and the explicit
construction of the magnetic amplitudes requires the use of the
locally defined potentials fulfilling $\eqref{gerb}$.

Let us now assume that a group $G$ acts on ${\cal M}$ and leaves $H$
as well as the kinetic term invariant. We aim at constructing magnetic
amplitudes for  some {\it open strings} in ${\cal M}$ that will define
{\it closed strings} on the orbifold ${\cal M}/G$. These are the
twisted sectors defined by strings that close up to an element of the
group. More precisely, the space of strings with winding $w$ is
\be
\cC_{w}=\la\,X_{w}:\,[0,2\pi]\rightarrow\cM\quad\mbox{such that}\quad
X(2\pi)=X(0)\cdot w\ra.
\ee
For $w=e$ the identity of $G$, it corresponds to closed strings and
one can glue together the endpoints of the segment to recover the
circle $S^{1}$. In the general case, it is convenient to view strings
in $\cC_{w}$ as defined on a pointed circle, with boundary values
differing by $w$, or equivalently, as strings defined on the universal
cover $\dR$ of $S^{1}$ fulfilling the quasi periodicity condition
$X_{w}(\sigma+2n\pi)=X_{w}(\sigma)\cdot w^{n}$. In
order to make the illustrations easier, we shall adopt the first
point of view.

The simplest magnetic amplitude corresponds to a single free string of
winding $w$
propagating between an initial position $X_{w}$ and a final one $Y_{w}$.
We construct the amplitude using a slight modification
of the powerful techniques introduced in \cite{gawedzki} (with the sign
conventions of \cite{newgawedzki}), in order to deal with
the twisted sectors. Such an amplitude is associated to a map $\varphi$
from a cut cylinder $\cS_{c}$ to $\cM$, that interpolates
between $X_{w}$ and $Y_{w}$ such that its boundary values along the cut
differ by $w$.
\begin{figure}[h]\centering
\includegraphics[width=9cm]{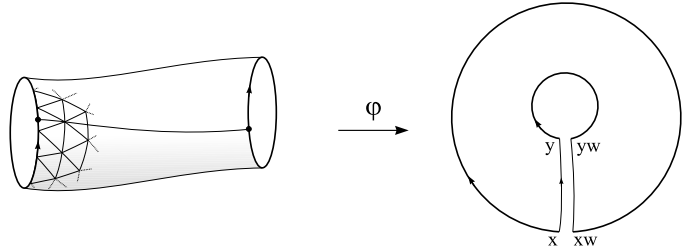}
\caption{The cut of a cylinder.}
\label{_cylinder}
\end{figure}
%


Pick up a triangulation of the cut cylinder by 2-simplices
$s_{\alpha}$, 1-simplices $l_{\beta}$ and vertices $v_{\gamma}$ such
that there exists an assignment of open sets fulfilling
$\varphi(s_{\alpha})\subset U_{i_{\alpha}}$, $\varphi(l_{\beta})\subset
U_{i_{\beta}}$ and $\varphi(v_{\gamma})\in U_{i_{\gamma}}$.
The triangulation  also includes an indepedent  triangulation of the cut by
segments $l'_{\beta'}$ and vertices $v'_{\gamma'}$ which we assume to
be identical on the two lips of the cut. That such
a triangulation exists follows from the invariance property of the
cover $\la U_{i}\ra$.

We define the corresponding magnetic amplitude as
\begin{eqnarray}
{\cal A}[\varphi]&=&
\prod\limits_{s_{\alpha}}\exp \ii\la
\int_{s_{\alpha}}\varphi^{\ast}B_{i_{\alpha}}\ra
\prod
\limits_{\begin{subarray}{c} s_\alpha\\ l_{\beta}\subset \partial s_{\alpha}\end{subarray}}
\exp \ii\la
\int_{l_{\beta}}\varphi^{\ast}B_{i_{\alpha}j_{\beta}}\ra
\prod
\limits_{\begin{subarray}{c} s_\alpha \\ l_{\beta}\subset\partial s_{\alpha}\\
v_{\gamma}\in l_{\beta} \end{subarray} }
\!\!\!\la
\varphi^{\ast}f_{i_{\alpha}j_{\beta}k_{\gamma}}^{\epsilon_{\beta\gamma}}(v_{\gamma})\ra \nonumber \\
&\times&\prod\limits_{l'_{\beta'}}\exp \ii\la
\int_{l'_{\beta'}}\varphi^{\ast}A_{w;\,j_{\beta'}}\ra
\prod\limits_{\begin{subarray}{c}
 l'_{\beta'} \\ v'_{\gamma'}\in\partial l'_{\beta'} \end{subarray}}
\!\!\!\la
\varphi^{\ast}f_{w;\,j_{\beta'}k_{\gamma'}}^{-\epsilon_{\beta'\gamma'}}(v'_{\gamma'})\ra
\label{fullamplitude}
\end{eqnarray}
The first line is identical to the contribution of closed strings in
\cite{gawedzki} and only involves the fields contained in ${\bf
  B}$. The second line is formally identical to the magnetic
amplitude for a particle (\ref{holo}), but remember that $f_{w; ij}$
are not the transition functions of a line bundle because of the third
equation in \eqref{cech}. This is reminiscent of the open string
amplitudes \cite{newgawedzki}, but ${\bf A}_{w}$ is not a new field of
the theory since it is related to ${\bf B}$ by
$\cD{\bf A}_{w}=w^{\ast}{\bf B}-{\bf B}$.
The rationale for this construction is to
compensate all the defects of the open string amplitude on the cut by the
contribution of the fields included in ${\bf A}_{g}$ integrated
along the cut.

To compare the amplitudes associated to two different triangulations
and assignments coinciding on the boundary strings $X_{w}$ and
$Y_{w}$, one proceeds as follows. First, let us notice that
it is sufficient to compare a triangulation to a finer one, i.e. a
second triangulation that contains all the simplices of the first
one. Then, we first compare all the terms arising form the 2-form
$B_{i}$ in ${\bf B}=(B_{i},B_{ij},f_{ijk})$. Their integrals agree up
to a discrepancy that only involves the 1-forms $B_{ij}$. The
contribution of the latter also cancels with the scalars $f_{ijk}$,
provided the corresponding edges do not meet the boundary. Next,
compare all terms on the two lips of the cut, taking into account the
contributions of the fields in ${\bf
  A}_{w}=(A_{w;\,i},f_{w;\,ij})$. Using \eqref{cech}, it turns out
that all the terms on the upper lip cancel with all the terms on the
lower lip, except on the boundary of the cut. This is a general rule:
In proving any identity, one should
always first compare the contribution of the top degree terms and add the
corresponding discrepancy to the lower degree terms. Then, one repeats the
procedure down to degree 0. Because of the high number of \Cech{}
indices involved in any non trivial computation, we do not display
them here. Some typical examples are worked out in detail in \cite{thesis}.

Accordingly, the amplitude only depends on the fields ${\bf B}$ and
${\bf A}_{w}$ as well as on the triangulations and assignments
$I$ and $J$ pertaining to the boundary strings $X_{w}$ and $Y_{w}$.
Because the expression
\eqref{fullamplitude} is rather
cumbersome to work with, we abbreviate it as
\be
\cA_{JI}[\varphi]=\lb \mathrm{e}^{\ii\int_{\Sigma}{\bf B}+
\ii\int_{x}^{y}{\bf A}_{w}}\rb_{JI},\label{cylamp}
\ee
where $x=X(0)$, $y=Y(0)$ and $\Sigma=\varphi({\cal S}_c)$. It really
means that one has to cut the
cylinder, triangulate it and compute the amplitude as in
(\ref{fullamplitude}). For topologically trivial $B$-fields, ${\bf B}$ and
${\bf A}$ can be identified with de Rham forms and (\ref{cylamp}) can be
taken literally as the integrals of ${\bf B}$ over $\Sigma$ and of ${\bf A}_{w}$
over the cut joining $x$ to $y$. In any case, the contributions of ${\bf B}$
and ${\bf A}_{w}$ cannot be separated, only their combination is
consistent.

The dependence of the amplitude on $I$ and $J$ is not very surprising.
Indeed, recall that in the case of a particle, the amplitude is to be
considered as a map from the fiber at $x$ to the fiber at $y$ of the
bundle whose sections are the wave functions. Therefore, its
definition involves a choice of open sets to cover $x$ and $y$.
Something similar happens for strings. $I$ and $J$ are indices that
label a covering of $\cC_{w}$ by open sets $U_{I}$ and the string wave
functions are sections of a bundle over $\cC_{w}$, defined using
transition functions $G_{IJ}$. The construction of the open sets
covering $\cC_{w}$ and of the transition functions also follows from
\cite{gawedzki}.

If $I$ is a triangulation of the cut circle by segments $l_{\alpha}$ and
vertices $v_{\beta}$ and an assignment of open sets $U_{i_{\alpha}}$
and $U_{j_{\beta}}$ that agree on the endpoints, we define
\be
U_{I}=\la\,X\in\cC_{w}\quad\mbox{such that}\quad
X(l_{\alpha})\subset U_{i_{\alpha}}
\quad\mbox{and}\quad X(v_{\beta})\in U_{i_{\beta}}\ra.
\ee
Varying over all triangulations and assignments, these open sets cover
$\cC_{w}$.

%
Consider now an other triangulation $J$ of the cut circle by
segments $l'_{\alpha'}$ and vertices $v'_{\beta'}$, and an assignment
of open sets $U_{i'_{\alpha'}}$ and $U_{j'_{\beta'}}$.
If $U_I\cap U_J$ is non empty, define a new triangulation of the intersection
by segments $\bar{l}_{\bar{\alpha}}:=l_\alpha\cap l'_{\alpha'}$ and associated vertices
$\bar{v}$.
Further, set $i_{\bar{\alpha}}=i_\alpha$ and
$i'_{\bar{\alpha}}=i'_{\alpha'}$. For $\bar{v}$ set
\begin{equation}
j_{\bar{\beta}}=\left\{ \begin{array}{c} j_\beta \\ i_\alpha \end{array} \quad\mathrm{if}\quad \begin{array}{c} \bar{v}=v_\beta \\ \bar{v}\in l_\alpha \end{array}\right.\quad,\quad
j'_{\bar{\beta}}=\left\{ \begin{array}{c} j'_{\beta'} \\ i'_{\alpha'} \end{array} \quad\mathrm{if}\quad \begin{array}{c} \bar{v}=v'_{\beta'} \\ \bar{v}\in l'_{\alpha'} \end{array}\right..
\end{equation}
Then, the transition functions $G_{IJ}$ are defined as
\begin{equation}
G_{IJ}(X)=
\prod\limits_{\bar{l}}
\exp\ii\left\{
\int_{\bar{l}} X^{\ast}\,B_{i_{\bar{\alpha}}i'_{\bar{\alpha}}}\right\}
\prod\limits_{\bar{v},\bar{l}|\bar{v}\in\partial\bar{l}}
\left(
\frac{ X^{\ast}f_{j_{\bar{\beta}}j'_{\bar{\beta}}i'_{\bar{\alpha}}}(\bar{v}) }
{X^{\ast}f_{j_{\bar{\beta}}i_{\bar{\alpha}}i'_{\bar{\alpha}}}(\bar{v})}
\right)^{-\epsilon_{\bar{\beta}}}
\times X^{\ast}f^{-1}_{w;\,j_{\beta_0}j'_{\beta'_0}}(0)
\label{transition}
\end{equation}
with $j_{\beta_0}$ and $j'_{\beta'_0}$ being the indices of open sets covering the cut
and $\epsilon_{\bar{\beta}}$ takes values $-1$ if
$\bar{v}$ is the first vertex of $\bar{l}$, and $+1$ if $\bar{v}$ is the second.
%
When three opens sets intersect, the consistency condition for the
transition functions read
\be
G_{IK}(X)G_{KJ}(X)=G_{IJ}(X),
\ee
for $X\in U_{I}\cap U_{J}\cap U_{K}$. To check this identity, one has
to use the three triangulations of the pointed circle and first compare the
contributions of the 1-forms on both sides. They agree up to a
boundary term, that is needed in order to cancel the discrepancy at the
cut. The latter arises because  $f_{w;\,ij}$ fails to be the transition
function of a bundle (see equation \eqref{cech}).

Accordingly, a line bundle can be constructed on $\cC_{w}$ using these
transition functions. The string wavefunction $\Psi$ is a section of this
bundle defined locally in the chart $U_{I}$ by a complex valued
function $\Psi_{I}$ that fulfills $\Psi_{I}=G_{IJ}\Psi_{J}$ on
overlapping charts. We denote by $\cH_{w}$ the space of sections of
the line bundle over $\cC_{w}$, it is the Hilbert space of strings
with winding $w$. Note that when the winding is trivial ($w=e$),
the previous construction reduces to the one presented in \cite{gawedzki}.

The dependence of the magnetic amplitudes on the covering and
assignments pertaining to the endpoints is readily expressed using
the transition functions. Indeed, if one trades
$I$ and $J$ for $K$ and $L$, a simple calculation shows that
\be
\lb \mathrm{e}^{\ii\int_{\Sigma}{\bf B}+\ii\int_{x}^{y}{\bf A}_{w}}\rb_{LK}
=
G_{LJ}(Y)\,
\lb \mathrm{e}^{\ii\int_{\Sigma}{\bf B}+\ii\int_{x}^{y}{\bf A}_{w}}\rb_{JI}\,G_{IK}(X).\;
\ee
This is in agreement with its interpretation as an holonomy in the line
bundle. Besides, the line bundle admits a connection that can be constructed
along the lines of \cite{gawedzki}. Roughly speaking, it corresponds
to the infinitesimal version of the holonomy, but its precise
mathematical definition is more involved because of the infinite dimensional
nature of $\cC_{w}$.\\

Consider now gauge transformations defined  by
$\Lambda=\lp \Lambda_{i},\eta_{ij}\rp$ and
$\Theta_{g}=\lp\theta_{g;i}\rp$, whose explicit actions on the fields
are given by  \eqref{gaugestring1} for $\Lambda$ and \eqref{gaugestring4} for
$\Theta$. This induces a change in the
amplitude
\be
\lb \mathrm{e}^{\ii\int_{\Sigma}{\bf B}+\ii\int_{x}^{y}{\bf A}_{w}}\rb_{JI}
\rightarrow
\lb \mathrm{e}^{\ii\int_{\Sigma}\lp {\bf B}+\cD\Lambda\rp+\ii\int_{x}^{y}\lp {\bf A}_{w}+
w^{\ast}\Lambda-\Lambda+\cD\Theta_{w}\rp}\rb_{JI}.
\label{gaugePsi}
\ee
Upon expressing everything using the triangulations and the explicit
form of the gauge transformations, we see that
there are cancelations that occur simplex by simplex in such a way
that only boundary terms remain.  The contribution along the cut of $\Lambda$
cancels with the lateral boundary of the worlsheet $\Sigma$, only the
integrals of $\Lambda$ along the incoming and outcoming string remain,
together with $\Theta_{w}$ and its inverse evaluated at $y$ and
$x$. Therefore, the gauge transformed amplitude reads
\be
\lb\Theta^{-1}_{w}(y)\mathrm{e}^{\ii\int_{y}^{yw}\Lambda}\rb_{J}
\lb\mathrm{e}^{\ii\int_{\Sigma}{\bf B}+\ii\int_{x}^{y}{\bf A}_{w}}\rb_{JI}
\lb \Theta_{w}(x)\mathrm{e}^{-\ii\int_{x}^{xw}\Lambda}\rb_{I},
\ee
where $\lb\cdots\rb_{I}$ always means that the expression inside the
bracket is evaluated using the triangulation and assignment given by
$I$.

For physics to remain invariant, the wave function has to be changed as
\be
\Psi_{I}(X)\rightarrow
\lb\Theta_{w}^{-1}(x)\mathrm{e}^{\ii\int_{x}^{xw}\Lambda}\rb_{I}\Psi_{I}(X).
\ee
It is worthwhile to notice that in addition to the expected 1-form
gauge transformation given by $\Lambda$, there is a new, winding
dependent scalar gauge tranformation associated to $\Theta_{w}$. In
the sequel, we shall refer to the latter as {\it secondary gauge transformations}.

\begin{figure}[h]\centering
\includegraphics[width=7cm]{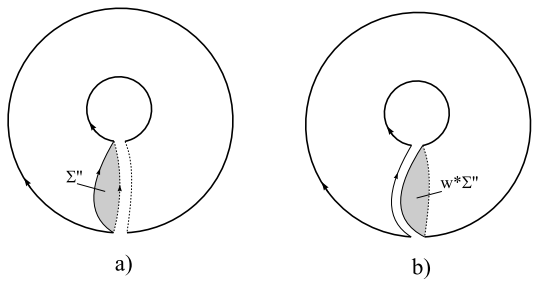}
\caption{Difference between $\Sigma$ and $\Sigma'$}
\label{_homotopy}
\end{figure}

The amplitude is also invariant under homotopic changes of the
cut $c$ with endpoints fixed. To show this, we first extend the map
$\varphi$ to the universal cover $\widetilde{\cal S}$ of the cylinder,
taken as a strip, by quasi periodicity. Because we assume that the
cover is invariant under the action of $G$, we can also extend it to a
periodic triangulation of $\widetilde{\cal S}$. Any other cut $c'$ defines
another map $\varphi'$ that coincides with $\varphi$ up to the action of
elements of $G$ of the type $w^{n}$, $n\in{\Bbb Z}$, on $\varphi$ on various
regions of $\widetilde{\cal S}$. Then we refine the triangulation
used for $\varphi$ in order that it covers the new cut, leaving it
unchanged on the boundary strings. Then, the new triangulation can
also be used to compute the magnetic amplitude for $\varphi'$ and the
equality of the two amplitudes follows from repeated use of the
relations \eqref{cech} to compare the integrals over the regions
differing by the action of $w^{n}$.
This is illustrated in figure \ref{_homotopy}, where the change of the
cut amounts to shifting the shaded area $\Sigma''$ by $w$, with
$\Sigma''$ corresponding to the difference between the surfaces $\Sigma$
and $\Sigma'$ provided by the maps  $\varphi$ and $\varphi'$.
\begin{figure}[t]\centering
\includegraphics[width=9cm]{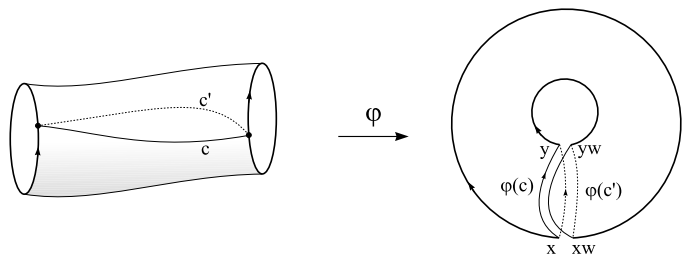}
\caption{Homotopic  change of the cut.}
\label{_cuts}
\end{figure}
%
%

Besides, the amplitude is invariant under diffeomorphisms of the
cylinder that are connected to the identity and reduce to the identity
on the boundary. Indeed, the latter simply
reduce to a change of the triangulation followed by a change of variables
in the associated integrals, as well as a smooth deformation of the
cut. This is not the case for large diffeomorphisms because the
latter induce a non homotopic change of the cut, as we shall see in
section \ref{MCG}.

\subsection{Stringy magnetic translations}

At the classical level, the group $G$ transforms a string $X_{w}$ that starts
at $x$ and ends at $x\!\cdot\! w$, into the string $X_{w}\!\cdot\!g$,
starting at $x\!\cdot\!g$ and ending
at $x\!\cdot\!wg=x\!\cdot\!g\,w^{g}$, with $w^{g}=g^{-1}wg$. Accordingly,
$g$ changes the winding and defines a map from $\cC_{w}$ to $\cC_{w^{g}}$.

In analogy with the particle's case, the stringy magnetic translation
$T_{g}^{w}:\,{\cal H}_{w^{g}}\rightarrow{\cal H}_{w}$
realizing this operation at the quantum level is the pullback action
on sections of the corresponding bundles, followed by a multiplication
by a winding dependent phase factor,
\be
\lb T_{g}^{w}\Psi\rb_{I}(X_{w})=
\lb\Upsilon_{g}^{w}(X_{w})\rb_{I}\,\Psi_{I}(X_{w}\!\cdot\!g),
\ee
for any $X_{w}\in U_{I}\subset\cC_{w}$. Because the open cover of $\cM$ is
invariant, we are allowed to use $U_{I}$ to cover $\cC_{w^{g}}$, with
the same triangulation and assignment. This can be summarized by the
assertion that the collection of all $U_{I}$ forms an invariant cover of
$\cC=\cup_{w} \cC_{w}$. Once we will have determined the
covering dependent phase $\lb\Upsilon_{g}^{w}(X)\rb_{I}$, it will be
necessary to show that all $\lb T_{g}^{w}\Psi\rb_{I}$ can be glued
together to form a section over $\cC_{w}$ that belongs to
$\cH_{w}$. Accordingly, $T_{g}^{w}$ will be a linear map from
$\cH_{w^{g}}$ into $\cH_{w}$.

As for the particle, we determine the unknown phase by requiring that
$T_{g}^{w}$ commutes with string propagation. To obtain a
sufficient condition ensuring this commutation, let us perform a
heuristic analysis, similar to the path integral derivation in section
\ref{particleproj}. The  propagation process of a string of winding $w$
is expressed as
\be
\Psi_{w}(X)\rightarrow \Psi'_{w}(Y)=
\int [DX]\,K_{w}(Y,X)\Psi_{w}(X),
\ee
where we have omitted the index $I$ for clarity but have displayed the
windings. In terms of matrix elements, the commutation relation
$T_g^w K_{w^g} = K_w T_g^w$ reads
\be
\Upsilon_{g}^{w}(Y)
K_{w^{g}}(Y\!\cdot\!g,X\!\cdot\!g)=
K_{w}(Y,X)\Upsilon_{g}^{w}(X),\label{commutation}
\ee
which is analogous to \eqref{propcom}.

Assuming that all terms in the path integral \eqref{stringpath}  but
the magnetic amplitudes are genuinely invariant under the action of
$G$ (as is the case for WZW models), it is sufficient to analyse the
behavior of the latter under the action of $G$. By equating the
magnetic contribution of $\varphi$ and $\varphi\cdot g$ in the path
integral on both sides of \eqref{commutation}, \be
\lb\Upsilon_{g}^{w}(Y)\rb_{J}\cA_{JI}[\varphi\!\cdot\!g]=
\cA_{JI}[\varphi]\lb\Upsilon_{g}^{w}(X)\rb_{I},\label{sufficient}
\ee we get a sufficient condition for the commutation to hold.

The expression of the magnetic amplitude for $\varphi\!\cdot\!g$ follows
from the results of the previous section,
\be
\cA_{JI}[\varphi\!\cdot\!g]=
\lb \mathrm{e}^{\ii\int_{\Sigma}g^{\ast}{\bf B}+\ii\int_{x}^{y}g^{\ast}{\bf A}_{w^{g}}}\rb_{JI}.
\ee
Recall that $g^{\ast}{\bf B}$ and $g^{\ast}{\bf A}_{w^{g}}$ are related to ${\bf B}$
and ${\bf A}_{w^{g}}$ by
\eqref{groupexplicit}, which implies
\be
\la
\begin{array}{rcl}
g^{\ast}{\bf B}&=&{\bf B}+\cD {\bf A}_{g},\\
g^{\ast}{\bf A}_{w^{g}}&=&{\bf A}_{wg}-{\bf A}_{g}+\cD\Phi_{g,w^g}.
\end{array}
\right.
\ee
To eliminate the unwanted ${\bf A}_{w^g}$, it is useful to introduce
\be
\Gamma_{w,g}=\Phi_{g,w^{g}}\lp\Phi_{w,g}\rp^{-1},\label{defGamma}
\ee
which fulfills
\be
\cD\Gamma_{w,g} + w^\ast{\bf A}_{g} - {\bf A}_g = g^\ast{\bf A}_{w^g}-{\bf A}_{w}.
\ee
Therefore, the ratio of the two amplitudes reads
\be
\frac{\cA_{JI}[\varphi\!\cdot\!g]}{\cA_{JI}[\varphi]}=
\lb \mathrm{e}^{\ii\int_{\Sigma}\cD {\bf A}_{g}+\ii\int_{x}^{y}
 \lp\cD\Gamma_{w,g}+w^{\ast}{\bf A}_{g}-{\bf A}_{g}\rp}\rb_{JI}.
\ee
This expression can be calculated using a triangulation and an
assignment of open sets, coinciding with $I$ and $J$ on the boundary
and expressing ${\bf A}_{g}$ and $\Gamma_{w,g}$ as \Cech{} cochains.
When performing this computation, lots of cancelations  occur and only
boundary terms remain. The final result is
\be
\frac{\cA_{JI}[\varphi\!\cdot\!g]}{\cA_{JI}[\varphi]}=
\lb \Gamma_{w,g}^{-1}(y)\mathrm{e}^{\ii\int_{y}^{yw}{\bf A}_{g}}\rb_{J}
\lb \Gamma_{w,g}(x)\mathrm{e}^{-\ii\int_{x}^{xw}{\bf A}_{g}}\rb_{I}.
\label{ratio}
\ee
Because of its cumbersome nature, the computation cannot be displayed
here, but it is instructive to check it for topologically trivial
fields. In this case, ${\bf A}_{g}$ is an ordinary 1-form and $\cD$ is the de
Rham differential, so that Stockes theorem yields
\be
\int_{\Sigma}\d A_{g}=\int_{\partial\Sigma}A_{g}.
\ee
The integral over the boundary is expressed as
\be
\int_{\partial\Sigma}A_{g}=\int_{y}^{yw}A_{g}-\int_{x}^{xw}A_{g}+
\int_{x}^{y}\lp A_{g}-w^{\ast}A_{g}\rp.
\ee
The integral over the cut cancels with the corresponding term in
\be
\int_{x}^{y}\lp\ii\d\log\Gamma_{w,g}+w^{\ast}A_{g}-A_{g}\rp,
\ee
so that we are left with boundary terms only, in accordance with \eqref{ratio}.

Turning back to the general case, we compare \eqref{ratio} with
\eqref{sufficient} and set
\be
\lb \Upsilon_{g}^{w}(X)\rb_{I}=
\lb\Gamma_{w,g}(x)\,\mathrm{e}^{-\ii\int_{x}^{xw}{\bf A}_{g}}\rb_{I}.
\ee
Accordingly,  the stringy magnetic translations are defined as
\be
\lb T_{g}^{w}\Psi\rb_{I}(X)=
\lb\Gamma_{w,g}(x)\,\mathrm{e}^{-\ii\int_{x}^{xw}{\bf A}_{g}}\rb_{I}\,
\Psi_{I}(X\!\cdot\!g),
\label{definition}
\ee
and the discussion preceding \eqref{sufficient} ensures that it
commutes with propagation.

Strictly speaking, we have defined $T_{g}^{w}$ only for wave functions defined
on local charts. To have operators defined between $\cH_{w^{g}}$ and
$\cH_{w}$, we have to check that all $\lb T_{g}^{w}\Psi\rb_{I}$ for
various $I$ can be glued into a global section on $\cC_{w}$. This
follows from the relation
\be
\frac{G_{IJ}(X\!\cdot\!g)}{G_{IJ}(X)}=
\frac{\lb\Gamma_{w,g}(x)\,\mathrm{e}^{-\ii\int_{x}^{xw}{\bf A}_{g}}\rb_{I}}
{\lb\Gamma_{w,g}(x)\,\mathrm{e}^{-\ii\int_{x}^{xw}{\bf A}_{g}}\rb_{J}},
\ee
that is obtained from the explicit form of $G_{IJ}$ given in
\eqref{transition}.
\medskip

\begin{figure}[h]\centering
\includegraphics[width=2cm]{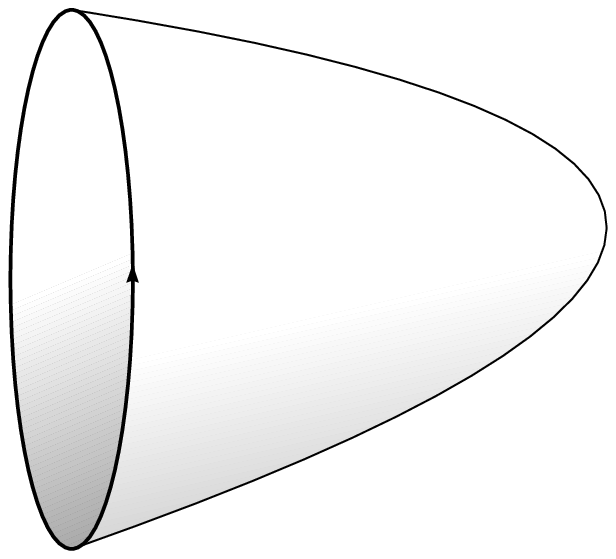}
\caption{A cap.}
\label{_cap}
\end{figure}

There is a very simple illustration of the action of a stringy
magnetic translation. Consider the process in which a single string with
trivial winding $w=e$ is created out of the vacuum. The corresponding
magnetic amplitude is
\be
\cA_{I}[\varphi]=\lb \mathrm{e}^{\ii\int_{\Sigma}{\bf B}}\rb_{I},
\ee
where $\Sigma=\varphi({\cal S})$ is a cap as in figure \ref{_cap}.
This amplitude is to be inserted in a path integral \eqref{stringpath},
\begin{equation}
\Psi_{I}(X)=\mathop{\int[D\varphi]}\limits_{\varphi(\partial{\cal S})=X}
\,e^{-S[\varphi]}{\cal A}_{I}[\varphi],
\end{equation}
and yields a wave function for a string of trivial winding, which is
obviously invariant under $G$.

\subsection{Multiplication law}

\label{orbifold}

In the particle's case, we have seen that magnetic translations only
form a projective representation of $G$. In order to check if
something similar happens for a string, let us compare
$T_{g}^{w}T_{h}^{w^{g}}$ with $T_{gh}^{w}$. On the left hand side, starting
with $\Psi\in\cH_{w^{gh}}$, we first act with $T_{h}^{w^{g}}$
\be
\Psi'_{I}(X)=\lb T_{h}^{w^{g}}\Psi\rb_{I}(X)=
\lb\Gamma_{w^{g},h}(x)\,\mathrm{e}^{-\ii\int_{x}^{xw^{g}}{\bf A}_{h}}\rb_{I}\,
\Psi_{I}(X\!\cdot\!h), \quad\mathrm{with}\,X\in{\cal C}_{w^{g}}.
\ee
Then, we further act on $\Psi'\in{\cal H}_{w^{g}}$ with $T_{g}^{w}$,
\be
\lb T_{g}^{w}\Psi'\rb_{I}(X)=
\lb\Gamma_{w,g}(x)\,\mathrm{e}^{-\ii\int_{x}^{xw}{\bf A}_{g}}\rb_{I}\,
\Psi'_{I}(X\!\cdot\!g),\quad\mathrm{with}\,X\in{\cal C}_{w}.
\ee
Replacing $\Psi'$ by its expression, we arrive at
\be
\lb T_{g}^{w}T_{h}^{w^{g}}\Psi\rb_{I}(X)=
\lb\Gamma_{w,g}(x)\,g^{\ast}\Gamma_{w^{g},h}(x)
\,\mathrm{e}^{-\ii\int_{x}^{xw}\lp {\bf A}_{g}+g^{\ast}{\bf A}_{h}\rp}\rb_{I}\,
\Psi_{I}(X\!\cdot\!gh),\quad\mathrm{with}\,X\in{\cal C}_{w},
\ee
where we have used the pullback operation of $g$ to trade the
integral of ${\bf A}_{h}$ over the string $X\!\cdot\!g$ for the integral of
$g^{\ast}{\bf A}_{h}$ over $X$.

\begin{figure}[h]\centering
\includegraphics[width=8cm]{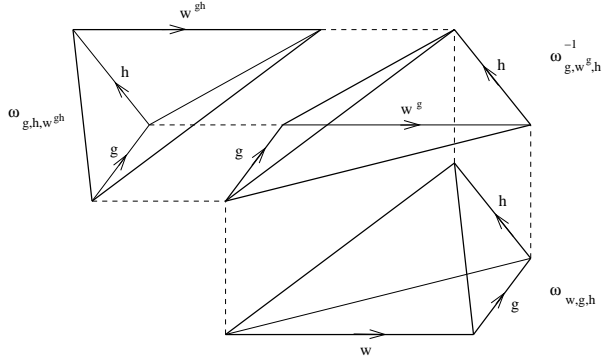}
\caption{The phase of the product in terms of tetrahedra.}
\label{_prism1}
\end{figure}

On the other hand side, we act directly with $T_{gh}^{w}$ to get
\be
\lb T_{gh}^{w}\Psi\rb_{I}(X)=
\lb\Gamma_{w,gh}(x)\,\mathrm{e}^{-\ii\int_{x}^{xw}{\bf A}_{gh}}\rb_{I}\,
\Psi_{I}(X\!\cdot\!gh).
\ee
Thus, there is a phase mismatch between $T_{gh}^{w}$ and
$T_{g}^{w}T_{h}^{w^{g}}$,
\be
T_{g}^{w}T_{h}^{w^{g}}=\frac{\lb\Gamma_{w,g}(x)\,g^{\ast}\Gamma_{w^{g},h}(x)
\,\mathrm{e}^{-\ii\int_{x}^{xw}\lp {\bf A}_{g}+g^{\ast}{\bf A}_{h}\rp}\rb_{I}}
{\lb\Gamma_{w,gh}(x)\,\mathrm{e}^{-\ii\int_{x}^{xw}{\bf A}_{gh}}\rb_{I}}
\,T_{gh}^{w}.
\ee
Using $\cD\Phi_{g,h}=g^{\ast}{\bf A}_{h}-{\bf A}_{gh}+{\bf A}_{g}$, cancelations similar
to the ones in \eqref{ratio} occur in the explicit evaluation using
\Cech{} cochains. The result can be expressed solely with $\Phi$:
\be
\frac{\lb\Gamma_{w,g}(x)\,g^{\ast}\Gamma_{w^{g},h}(x)
\,\mathrm{e}^{\ii\int_{x}^{xw}\lp {\bf A}_{g}+g^{\ast}{\bf A}_{h}\rp}\rb_{I}}
{\lb\Gamma_{w,gh}(x)\,\mathrm{e}^{\ii\int_{x}^{xw}{\bf A}_{gh}}\rb_{I}}
=
\lb\frac{\Phi_{g,w^{g}}\,g^{\ast}\Phi_{h,w^{gh}}\,\Phi_{w,gh}\,w^{\ast}\Phi_{g,h}}
{\Phi_{w,g}\,g^\ast\Phi_{w^{g},h}\,\Phi_{gh,w^{gh}}\,\Phi_{g,h}}(x)\rb_{I}.
\ee
After repeated use of $\omega=\delta\Phi$, the previous expression
simplifies to
\be
\lb\frac{\Phi_{g,w^{g}}\,g^{\ast}\Phi_{h,w^{gh}}\,\Phi_{w,gh}\,w^{\ast}\Phi_{g,h}}
{\Phi_{w,g}\,g^\ast\Phi_{w^{g},h}\,\Phi_{gh,w^{gh}}\,\Phi_{g,h}}(x)\rb_{I}=
\frac{\omega_{w,g,h}\,\omega_{g,h,w^{gh}}}{\omega_{g,w^{g},h}}.
\ee
Because $\cD\omega=0$, $\omega$ is globally defined (i.e. it does not
depend on the open set in which it has been computed) and
constant. This is why we dropped the $I$ and $X$ in the notation.
The occurrence of this combination of the 3-cocycle $\omega$ corresponds
to chopping the prism spanned by $w$, $w^{g}$ and $w^{gh}$ into three
tetrahedra, as can be seen in figure \ref{_prism1}.
%

Finally, the product law reads
\begin{equation}
T_{g}^{w}T^{v}_{h}=\delta_{v,w^{g}}\,\frac{\omega_{w,g,h}\,\omega_{g,h,w^{gh}}}{\omega_{g,w^{g},h}}\,
T_{gh}^{w},\label{product}
\end{equation}
where we have defined the product as zero when
the windings do not match. This is exactly the multiplication of the
quasi-quantum group $D_{w}[G]$ introduced in \cite{dijkgraaf}. Thus,
as operators on the full Hilbert space ${\cal H}=\oplus_{w}{\cal
H}_{w}$, the $T_{g}^{w}$ generate the quasi-quantum group
$D_{w}[G]$. Though a 3-cocycle usually indicates a breakdown of an
associativity law, the product of three stringy magnetic
translations remains associative, since they are operators in the
Hilbert space. This last property can be checked directly using the
cocycle properties of $\omega$. Nevertheless, we shall see in the
next section that there is failure of associativity when considering
the interaction of three strings.
\medskip

The expression of the phase factor appearing in the multiplication
law \eqref{product} in terms of the 3-cocycle $\omega$ is easily
interpreted using a transgression map similar to the one introduced
in
\cite{willerton}.
Let us denote by $(C,\delta)$ the standard
group cohomology complex  with values in $U(1)$ and by
$(\widetilde{C},\tilde{\delta})$ a new complex whose cochains are
group cochains with values in $U(1)$-valued functions over $G$. For
any $n$-cochain in $\widetilde{C}^{n}$ defined by the function
$w\mapsto \widetilde{c}_{g_{1},\dots,g_{n}}(w)$, we define the
twisted coboundary as
\begin{equation}
(\tilde{\delta} \tilde{c})_{g_{0},\dots,g_{n}}(w)=\tilde{c}_{g_{1},\dots,
g_{n}}(w^{g_{0}})\times\prod_{k=1}^{n}\left(\tilde{c}_{g_{0},\dots,g_{k-1}g_{k},\dots,g_{n}}(w)\right)^{(-1)^{k}}
\times\left(\tilde{c}_{g_{0},\dots, g_{n-1}}(w)\right)^{(-1)^{n+1}}.
\label{twisted}
\end{equation}
The transgression map ${\cal T}$ takes an $(n+1)$-cochain $c_{g_1,\dots,g_{n+1}}\in C^{n+1}$ to
an $n$-cochain in $\widetilde{C}$ by
\begin{equation}
\left[{\cal T}c\right]_{g_{1},\dots,g_{n}}(w)= \prod_{i=0}^{n}
\left(c_{g_{1},\dots,g_{i},w^{g_{1}\dots g_{i}},g_{i+1},\dots,g_{n}}
\right)^{(-1)^{n-i}}.
\end{equation}
For cocycles $\beta$, $\alpha$ and $\omega$ of orders 1,2 and 3, we
have
\begin{equation}
\left[{\cal T}\beta\right](w)=\beta_{w},\quad \left[{\cal
T}\alpha\right]_{g}(w)=\frac{\alpha_{g,w^g}}{\alpha_{w,g}},\quad
\left[{\cal T}
\omega\right]_{g,h}(w)=\frac{\omega_{w,g,h}\omega_{g,h,w^{gh}}}{\omega_{g,w^{g},h}}.
\label{exampletrans}
\end{equation}
The transgression map induces a morphism of complexes since it
fulfills ${\cal T}\circ\delta=\tilde{\delta}\circ{\cal T}$.
Therefore, the image of an $(n+1)$-cocycle in $C$ is an $n$-cocycle in
$\tilde{C}$. In particular, the 3-cocycle $\omega$ is transgressed
into a 2-cocycle on functions of the winding, which allows to
interpret the product law \eqref{product} of stringy magnetic translations as a
projective representation.
\medskip

Let us close this section by a brief comment on the orbifold
construction. Starting with a string theory on ${\cal M}$, the latter
consists in two steps. First, one introduces the twisted sectors, that
form the Hilbert space $\cH$. Then, one
projects the Hilbert space onto the subspace made out of vectors that are
invariant (up to a phase) under the action of $G$. In case the
3-cocycle $\omega$ is non trivial, there are no invariant states with
a non trivial winding.  This is similar to the particle's case, as we
have seen in section \ref{particle}. If there
is a phase mismatch between $T_{g}T_{h}$ and $T_{gh}$, no non trivial
invariant states can be constructed in ${\cal H}$, unless $\omega$ is cohomologically trivial in the sense of \cite{coste}. Indeed, the representations of $D_{\omega}[G]$ can be understood as direct sums of projective representations in each conjugacy classes \cite{maillard}. In the orbifold string theory, these conjugacy classes define the sectors and for cohomologically trivial $\omega$ the phases of the operators $T_{g}^{w}$ may be adjusted in such a way that there is no phase in the multiplication law.  However, as we shall see in the last section, global anomalies remain when $\omega\neq1$. Therefore, the
triviality of $\omega$ appears as an additional consistency
obstruction for the orbifold theory, as already noticed in
\cite{sharpe}, but cannot be derived solely from the invariance of the
3-form $H$ under the group.

On the geometrical side, $\omega$ can be
considered as an obstruction to pushing forward the gerb on 
${\cal M}$
defined by $\eqref{gerb}$  to a gerb on ${\cal M}/G$ when the action
of $G$ is free. This interpretation is in accordance  with the results
obtained in \cite{simply} in case of simply connected Lie groups
quotiented by finite subgroups.

Even if we assume that $\omega=1$, it is in general impossible to gauge
away the extra phases $\Upsilon_{g}^{w}$ simultaneously for all
$g\in G$. Indeed, this would require $\Upsilon_{g}^{w}$ to be
independent of $g$, which implies that so do  ${\bf A}_{g}$ and
$\Gamma_{w}^{g}$. This is in general not the case, as can be checked
on the simple example given by a constant $H$ in \eqref{globalfields}.

\section{Interacting strings}

\subsection{Quasi Hopf structure from interactions}

\label{interaction}

The cornerstone of the subsequent derivation of the quasi-quantum
group structure of the algebra generated by the stringy magnetic
translations resides in the geometrical nature of the interactions
of strings. Indeed, let us consider a process in which $m$ incoming
strings scatter to produce $n$ outgoing ones, which gives rise to an
operator $K^{m\rightarrow n}$ from ${\cal H}^{\otimes m}$ to ${\cal
H}^{\otimes n}$. In the wave function approach adopted here, the
matrix elements of $K^{m\rightarrow n}$ can be derived from a two
dimensional field theory as
\begin{equation}
K^{m\rightarrow
n}(Y,X)=\mathop{\int[D\varphi]}\limits_{\varphi(\partial {\cal
S})=X^{*}\cup Y}\,\mathrm{e}^{-S[\varphi]},
\end{equation}
where $\varphi: {\cal S}\rightarrow {\cal M}$ is the conformal field
describing the embedding of the strings into space-time. Of course,
one has to further integrate over the moduli of the Riemann surface
${\cal S}$ and sum over all possible genera. The surface ${\cal S}$
has a boundary made of $m+n$ circles, the n circles
$Y=(Y_{1},\dots,Y_{n})$ whose orientation agrees with that of
$\varphi(\partial{\cal S})$ correspond to the outgoing strings while
the $m$ others $X=(X_{1},\dots,X_{m})$ are the incoming ones.

In the previous sections, we have already dealt with simple examples
restricted to free strings: $K^{1\rightarrow 1}$ and
$K^{0\rightarrow 1}$, describing respectively the propagation and
the creation of a single string out of the vacuum. In the sequel, we
shall deal with interacting strings, restricting ourselves to the
simplest processes. The first example we treat in detail is the tree
level contribution to the decay process $K^{1\rightarrow 2}$
involving pairs of pants. This allows us to derive the coproduct of
$D_{\omega}[G]$ from the requirement of the commutativity of
$K^{1\rightarrow 2}$ with stringy magnetic translations. Then, we
show that the defining relations for the antipode of $D_{\omega}[G]$
arises from the orientation reversing operation that relates
$K^{0\rightarrow 2}$ and $K^{1\rightarrow 1}$. Other features of the
quasi-quantum group $D_{\omega}[G]$ have natural counterparts in
the theory of interacting strings. We illustrate this for the
associator, the braiding and the invariance under Drinfel'd twists,
respectively related to the failure of the associativity of the
tensor product of several strings, the exchange of two strings and
discrete torsion.

\subsection*{Derivation of the coproduct}

\begin{figure}[h]\centering
\includegraphics[width=8cm]{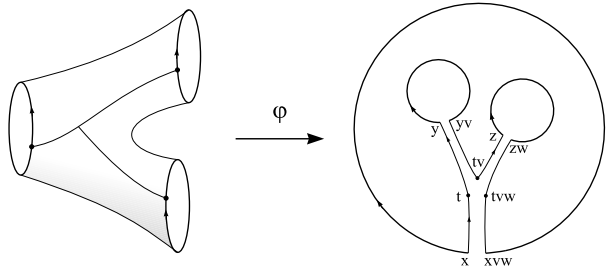}
\caption{Basic interaction associated to a pair of pants.}
\label{_hose}
\end{figure}

The most basic interaction is given by the tree level decay of a
string of winding $vw$ into a string of winding $v$ and another one
of winding $w$. The matrix element  $K^{1\rightarrow 2}(Y\cup Z,X)$
is associated to the pants ${\cal S}$ depicted in figure
\ref{_hose}.
%
To construct the corresponding magnetic amplitude for a map
$\varphi$ such that $\varphi(\partial{\cal S})=X^{*}\cup Y\cup Z$,
one has to cut the pants in order to obtain a simply connected
surface. Then, the magnetic amplitude reads \be
\cA_{JK,I}[\varphi]=\lb \mathrm{e}^{\ii\int_{\Sigma}{\bf B}
+\ii\int_{x}^{t}{\bf A}_{vw}+\ii\int_{t}^{y}{\bf
A}_{v}+\ii\int_{tv}^{z}{\bf A}_{w}}
\;\Phi_{v,w}^{-1}(t)\rb_{JK,I},\label{pant} \ee where $I$ is used to
cover the incoming string while $J$ and $K$ label the covers of the
two outgoing strings. The insertion of $\Phi^{-1}_{v,w}$ is
necessary to maintain invariance under the secondary gauge
transformations of ${\bf A}$ and $\Phi$ given by
\eqref{gaugestring3}. Indeed, 1-form gauge transformations of ${\bf
B}$ and ${\bf A}$ cancel together at the boundary $\partial\Sigma$,
except on the boundary strings as is expected. The secondary gauge
transformation induced by $\Theta\in C^{\mathrm{tot}}_{1,0}$ also
has boundary terms on the strings as expected by \eqref{gaugePsi},
and an extra contribution at the point $t$ where the three cuts
meet.  If we denote by $U_{l}$ the open set used to cover the point
$t$, a secondary gauge transformation yields
$[\Theta^{-1}_{vw}(x)]_{l}$ along the cut joining $x$ to $t$ and
$[\Theta_{v}(t)\Theta_{w}(t\!\cdot\!v)]_{l}$ along the two other
cuts leaving $t$. The net result is
$[v^{\ast}\Theta_{w}\Theta^{-1}_{vw}\Theta_{v}]_{l}$, which is
canceled by the secondary gauge transformation of
$[\Phi^{-1}_{v,w}(t)]$ given in \eqref{gaugestring3}. Obviously, if
we consider the reversed process in which two strings join, one has
to insert $\Phi_{v,w}$. It is interesting to note that a similar
phase factor has been encountered in the context of toroidally
compactified closed string field theory \cite{sft}. Whereas our
phase factor is a 2-cochain on the windings, their phase factor is a
2-cocycle on the momenta. This is equivalent, since T-duality
exchanges windings and momenta, and their cocycle condition simply
stipulates the triviality of $\omega$.

The amplitude is independent of the homotopy class of the three cuts
and under diffeomorphisms connected to the identity that act trivially
on the boundary strings, as one may check using the same techniques
as for the free string propagator. Besides, using the relation
${\cD}\Phi_{v,w}=v^{\ast}{\bf  A}_{w}-{\bf A}_{vw}+{\bf A}_{v}$,
one can also move the point $t$ along the cut, leaving the amplitude unchanged.

This amplitude contributes to the process describing the tree level
decay ${\cal H}_{vw}\rightarrow{\cal H}_{v}\otimes {\cal H}_{w}$. We
expect that the corresponding operator $K_{v,w}$
(the upper index $\cdot^{1\to 2}$ has been omitted for convenience)
commutes with the
action of $G$, but the latter may be defined on the tensor product
${\cal  H}_{v}\otimes{\cal H}_{w}$
only up to an additional  phase. Again, we determine this phase by an
heuristic analysis, focusing on the magnetic amplitude only.

The magnetic amplitude is to be inserted into the path integral for
the evolution operator $K_{v,w}$ encoding the decay process
$\cH_{vw}\rightarrow\cH_{v}\ot\cH_{w}$,
\be
K_{v,w}\lp Y\cup Z,X\rp=\mathop{\int
  [D\varphi]}\limits_{\varphi(\partial{\cal S})=X^{\ast}\cup Y\cup Z}
\,\mathrm{e}^{-S[\varphi]}\cA[\varphi],
\ee
where $X$, $Y$ and $Z$ are three strings of respective windings $vw$, $v$ and $w$.
In this heuristic analysis we have omitted indices relative to the coverings
for clarity.


To determine the possible extra phase in the tensor product, we
compare the operators $K_{v,w}T_{g}^{vw}$ and
$\lp T_{g}^{v}\otimes T_{g}^{w}\rp K_{v^{g},w^{g}}$. At the level of
the kernels, we have to compare
\begin{equation}
K_{v,w}(Y\cup Z,X)\Upsilon_{g}^{vw}(X)
\end{equation}
and
\begin{equation}
\Upsilon_{g}^{v}(Y)\Upsilon_{g}^{w}(Z)
K_{v^{g},w^{g}}(Y\!\cdot\!g\cup Z\!\cdot\!g,X\!\cdot\!g),
\end{equation}
with $\Upsilon_{g}^{w}$ the phase defining the stringy magnetic
translation.  Let us proceed as we did for single string propagations
in the last section by comparing the magnetic amplitudes.  ${\cal
  A}_{JK,I}[\varphi]$ enters into
$K_{v,w}(Y\cup Z,X)$ whereas ${\cal A}_{JK,I}[\varphi\!\cdot\!g]$ is the
corresponding term in
$K_{v^{g},w^{g}}(Y\!\cdot\!g \cup Z\!\cdot\! g,X\!\cdot\!g )$.

To begin with, the translated amplitude involving strings of windings
$v^{g}$, $w^{g}$ and $(vw)^{g}$ reads
\be
\cA_{JK,I}[\varphi\!\cdot\!g]=\lb \mathrm{e}^{\ii\int_{\Sigma}g^{\ast}{\bf B}
+\ii\int_{x}^{t}g^{\ast}{\bf A}_{(vw)^{g}}
+\ii\int_{t}^{y}g^{\ast}{\bf A}_{v^{g}}+\ii\int_{tv}^{z}g^{\ast}{\bf A}_{w^{g}}}
\;g^{\ast}\Phi^{-1}_{v^{g},w^{g}}(t)\rb_{JK,I}.
\ee
After some manipulations using the relations \eqref{cech},
the ratio can be expressed as
\be
\frac{\cA_{JK,I}[\varphi\!\cdot\!g]}{\cA_{JK,I}[\varphi]}
=\lb
\frac{\Gamma^{-1}_{v,g}(y)\,\mathrm{e}^{\ii\int_{y}^{yv}{\bf A}_{g}}\;
\Gamma^{-1}_{w,g}(z)\,\mathrm{e}^{\ii\int_{z}^{zw}{\bf A}_{g}}}
{\Gamma^{-1}_{vw,g}(x)\,\mathrm{e}^{\ii\int_{x}^{xvw}{\bf A}_{g}}}
\times\frac{\Phi_{v,w}\,\Gamma_{v,g}
\,v^{\ast}\Gamma_{w,g}}{g^{\ast}\Phi_{v^{g},w^{g}}\,\Gamma_{vw,g}}(t)
\rb_{JK,I}.\label{comparison}
\ee


The first ratio in the bracket can be written as
\begin{equation}
\frac{
\lb\Gamma^{-1}_{v,g}(y)\,\mathrm{e}^{\ii\int_{y}^{yv}{\bf A}_{g}}\rb_{J}\;
\lb\Gamma^{-1}_{w,g}(z)\,\mathrm{e}^{\ii\int_{z}^{zw}{\bf A}_{g}}\rb_{K}}
{\lb\Gamma^{-1}_{vw,g}(x)\,\mathrm{e}^{\ii\int_{x}^{xvw}{\bf A}_{g}}\rb_{I}}
=\frac{\lb\Upsilon_{g}^{vw}(X)\rb_{I}}
{\lb\Upsilon_{g}^{v}(Y)\rb_{J}\,\lb\Upsilon_{g}^{w}(Z)\rb_{K}}.
\end{equation}
It is exactly what one expects from
acting with $T_{g}^{vw}$ on the incoming string and $T_{g}^{v}\ot
T_{g}^{w}$ on the two outgoing ones, so that it cancels when acting
with the stringy magnetic translations.

The second one can be expressed using the 3-cocycle $\omega$ as
\be
\lb\frac{g^{\ast}\Phi_{v^{g},w^{g}}\,\Gamma_{vw,g}}{\Phi_{v,w}\,\Gamma_{v,g}
\,v^{\ast}\Gamma_{w,g}}(t)
\rb_{JK,I}
=
\frac{\omega_{v,w,g}\,\omega_{g,v^{g},w^{g}}}{\omega_{v,g,w^{g}}}.
\ee
Because its expression only involves $\omega$, it is globally defined and
constant.

Therefore, the action of $G$ on the tensor product
${\cal H}_{v^{g}}\otimes{\cal H}_{w^{g}}$ must be defined as
\begin{equation}
\frac{\omega_{v,w,g}\,\omega_{g,v^{g},w^{g}}}{\omega_{v,g,w^{g}}}\,
\lp T_{g}^{v}\otimes T_{g}^{w}\rp.
\end{equation}
This is nothing but the phase appearing in the coproduct
$\Delta:\,D_{\omega}[G]\rightarrow D_{\omega}[G]\otimes
D_{\omega}[G]$ of the quasi-quantum group, defined in
\cite{dijkgraaf} as
\begin{equation}
\Delta(T_{g}^{u})=\mathop{\sum}\limits_{vw=u}\,\frac{\omega_{v,w,g}\,\omega_{g,v^{g},w^{g}}}{\omega_{v,g,w^{g}}}\,
T_{g}^{v}\ot T_{g}^{w}.\label{coproduct}
\end{equation}
The occurrence of the coproduct in the action on the tensor product
is natural in the theory of quasi Hopf algebras (see appendix).
Indeed, given two representations  $\rho_{1}$ and $\rho_{2}$ of
$D_{\omega}[G]$, a new one can be build as
\begin{equation}
\lp\rho_{1}\otimes\rho_{2}\rp\circ\Delta.\label{tens}
\end{equation}
This is exactly what happens for strings in a background 3-form $H$:
the representation acting on ${\cal H}_{v}\otimes{\cal H}_{w}$ is
nothing but the tensor product of the representations on ${\cal H}_{v}$
and on ${\cal H}_{w}$, but with the tensor product defined using the
coproduct rule \eqref{tens}.

\begin{figure}[h]\centering
\includegraphics[width=9cm]{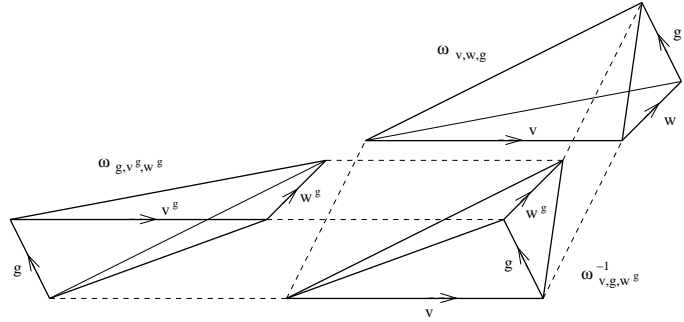}
\caption{The phase of the coproduct in terms of tetrahedra.}
\label{_prism2}
\end{figure}

As for the phase appearing in the product, the phase of the coproduct can
be understood in terms of a prism, representing the translation by
$g$ of a string of winding $vw$ decomposed into a string of winding $v$
followed by a string of winding $w$. The decomposition of the prism into the
three tetrahedra components is depicted in figure \ref{_prism2}.

This coproduct is only quasi-coassociative,
\begin{equation}
\lp \mathrm{id} \ot \Delta \rp \circ \Delta=
\Omega\big(\lp\Delta\ot\mathrm{id} \rp\circ\Delta\big)\Omega^{-1},
\label{quasi666}
\end{equation}
where $\Omega$ is the Drinfel'd associator, related to the 3-cocycle
$\omega$ by
\begin{equation}
\Omega=\mathop{\sum}\limits_{u,v,w}\omega_{u,v,w}^{-1}\,
T_{e}^{u}\ot T_{e}^{v}\ot T_{e}^{w}.
\end{equation}
Together with the antipode and the braiding to be defined in
subsequent sections, the quasi-quantum group $D_{\omega}$ turns out
to be a quasi-triangular quasi Hopf algebra.

In accordance with the general theory of quasi Hopf algebras, let us
note that $D_{\omega}[G]$ also comes equipped with a counit defined
as $\epsilon(T_{g}^{w})=\delta_{w,e}$ \cite{dijkgraaf}.

\medskip

In the case of globally defined fields given by \eqref{globalfields}
in the case $G={\Bbb Z}^{N}$, the product is \be T_{g}^{w}T_{h}^{v}=
\delta_{v,w}\;\mathrm{e}^{\,\frac{\ii}{6}H_{\mu\nu\lambda}w^{\mu}g^{\nu}h^{\lambda}}\;
T_{gh}^{w} \ee and the coproduct \be \Delta\lp
T_{g}^{w}\rp=\mathop{\sum}\limits_{uv=w}\,
\mathrm{e}^{\,\frac{\ii}{6}H_{\mu\nu\lambda}g^{\mu}u^{\nu}v^{\lambda}}\;
T_{g}^{u}\ot T_{g}^{v}. \ee As an algebra, it is nothing but a
direct sum of noncommutative tori with winding dependent phases
inserted in the multiplication law of the generators. The coproduct
mixes different windings and turns out to be coassociative because
the associator is central. From a pure mathematical viewpoint, this
is rather interesting because it is known that noncommutative tori
are not Hopf algebras. The Hopf algebra structure can only be
obtained by taking into account all the windings. However, the
associator does not drop from the quasi Yang-Baxter equation
\eqref{qybe} and the quasi triangular structure only holds in the
sense of quasi Hopf algebras.
\bigskip

As an aside, let us note that the product and the coproduct of
$D_{\omega}[G]$ admit a cohomological interpretation very similar to
the transgression already introduced in \eqref{exampletrans}.
Indeed, the consistency of the quasi Hopf algebra structure implies
that the phases appearing in the product \eqref{product}
\begin{equation}
{\cal
P}^{g,h}_{w}=\frac{\omega_{w,g,h}\,\omega_{g,h,w^{gh}}}{\omega_{g,w^{g},h}},
\end{equation}
and coproduct \eqref{coproduct}
\begin{equation}
{\cal Q}^{g}_{v,w}=
\frac{\omega_{v,w,g}\,\omega_{g,v^{g},w^{g}}}{\omega_{v,g,w^{g}}}
\end{equation}
obey the following equations
\begin{equation}
\left\{
\begin{array}{rcl}
{\cal P}_{u}^{g,h} \,{\cal P}_{u}^{gh,k} &=& {\cal P}_{u}^{g,hk}
\,{\cal P}_{u^{g}}^{h,k},\cr
{\cal P}_{u}^{g,h}\,{\cal
P}_{v}^{g,h}\,{\cal Q}^{g}_{u,v}\,{\cal Q}^{h}_{u^{g},v^{g}}&=&{\cal
P}_{uv}^{g,h}\,{\cal Q}^{gh}_{u,v}, \cr
{\cal Q}_{u,v}^{g}\,{\cal Q}_{uv,w}^{g}\,\omega_{u^{g},v^{g},w^{g}}&=&
{\cal Q}_{v,w}^{g}\,{\cal Q}_{u,vw}^{g}\,\omega_{u,v,w}.
\end{array}
\right.
\end{equation}
The first equation expresses the associativity of the product, the
second one the compatibility of the product and the coproduct and
the last one the quasi-coassociativity of the coproduct.

Using the group coboundary $\delta$ for the windings $u,v,w,\dots$
and the twisted group coboundary $\tilde{\delta}$ (generalizing the
one introduced in $\eqref{twisted}$) for $g,h,k,\dots$, with an
extra action by conjugation on the windings for its first factor,
these equations read
\begin{equation}
\left\{
\begin{array}{rcl}
(\tilde{\delta}{\cal P})_{u}^{g,h,k} &=& 1,\cr (\tilde{\delta}{\cal
Q})_{u,v}^{g,h}\, (\delta{\cal P})_{u,v}^{g,h}&=& 1,\cr (\delta{\cal
Q})_{u,v,w}^{g}\,
\left((\tilde{\delta}{\omega})_{u,v,w}^{g}\right)^{-1} &=& 1.
\end{array}
\right.
\end{equation}
Together with $(\delta\omega)_{u,v,w}=1$, these equations simply
stipulate that $({\cal P}^{g,h}_{u}, {\cal
Q}_{u,v}^{g},\omega_{u,v,w})$ is a 2-cocycle in a bicomplex
constructed out of $\delta$ and $\tilde{\delta}$.

\subsection*{Antipode}
In the general formula \eqref{interaction} encoding the scattering
process of several strings, whether a string is incoming or outgoing
depends on its orientation with respect to that of the surface
${\cal S}$ involved in the matrix element. However, a given ${\cal
S}$ can contribute to various processes differing solely by the
choice of the incoming and outgoing nature of the strings among the
circles defined by $\partial{\cal S}$.

This suggests the existence of an operator ${\Pi}$ from ${\cal H}$
to its dual that changes an outgoing string to an incoming one. At
the wave function level, if $\Psi\in{\cal H}_{w}$, then we define
$\Pi\Psi\in\left({{\cal H}_{w^{-1}}}\right)^{\ast}$ by
\begin{equation}
\left[\Pi\Psi(X^{\ast})\right]_{I^{\ast}}=\left[\Phi_{w,w^{-1}}(x)\right]_{I}
\left[\Psi(X)\right]_{I},
\end{equation}
where $X^{\ast}\in{\cal C}_{w^{-1}}$ denotes the string with
orientation reversed and $I^{\ast}$ is the triangulation and
assignment with order reversed. $\Pi\Psi$ is a section of the dual
bundle defined by the transition functions
$G_{IJ}^{\ast}=G_{IJ}^{-1}$ and with gauge transformations and
holonomies along cylinders defined by the opposite phases. As for
the pair of pants amplitude in the last section, the inclusion of
the winding dependent extra phase $\Phi_{w,w^{-1}}$ is dictated by
the requirement of invariance under secondary gauge transformations.
In geometrical terms, $\Pi$ is a line bundle isomorphism induces by
the orientation reversing.

For general Hopf algebras (see for instance \cite{chari}), the
antipode allows to define the dual of a given representation. In our
context, the antipode ensures the compatibility of the quasi-quantum
group action on the string states with the orientation reversing
operation. Indeed, if $\Psi\in{\cal H}_{w}$ and $\Psi'\in{\cal
H}_{(w^{-1})^{g}}$, then we have
\begin{equation}
\left[\Pi\left(
T_{g}^{w}\Psi\right)(X^{\ast})\right]_{I^{\ast}}\Psi'_{I}(X^{\ast})=
\left[\Pi\Psi(X^{\ast}\!\cdot\!g)\right]_{I^{\ast}}\left[S(T_{g}^{w})\Psi'(X^{\ast}\!\cdot\!g)\right]_{I}
\label{dualaction}
\end{equation}
for any string $X^{\ast}\in{\cal C}_{w^{-1}}$, with
\begin{equation}
S(T_{g}^{w})=
\frac{\omega_{g,(w^{-1})^{g},g^{-1}}}{\omega_{w^{-1},g,g^{-1}}\,\omega_{g,g^{-1},w^{-1}}}\,
\frac{\omega_{w,g,(w^{-1})^{g}}}{\omega_{w,w^{-1},g}\,\omega_{g,w^{g},(w^{-1})^{g}}}\,
T_{g^{-1}}^{(w^{-1})^{g}} \label{antipode2}
\end{equation}
the antipode of the quasi-quantum group (see \cite{dijkgraaf}). Note
that the first phase factor is nothing but the inverse of the phase
appearing in the product $T^{w^{-1}}_{g}T^{(w^{-1})^{g}}_{g^{-1}}$
(see \eqref{product}) while the second phase factor is the inverse
of the one appearing in the coproduct for a translation of $g$ of
two strings of windings $w$ and $w^{-1}$. Note that
\eqref{dualaction} can be used to derive the extra phase in
\eqref{antipode2}.

$S$ fulfills all the requirements imposed on the antipode of a
quasi-Hopf algebra (see appendix), with trivial $\alpha$ and
\begin{equation}
\beta=\sum_{w}\omega_{w,w^{-1},w}\,T_{g}^{w}.
\end{equation}
Besides, its square obeys
\begin{equation}
S^{2}(a)=\beta^{-1}a\beta\label{squareofs}
\end{equation}
for any $a\in D_{\omega}[G]$, as shown in \cite{coste}.

To illustrate the previous construction on a simple example, let us
consider a two string state  $K^{0\rightarrow 2}$ created out of the
vacuum. Reversing the orientation of the first string, we recover
the single string propagator
\begin{equation}
(\Pi\otimes\mbox{id})K^{0\rightarrow 2}=K^{1\rightarrow1}.
\end{equation}
At the level of matrix elements, this equation reads
\begin{equation}
\Phi_{w,w^{-1}}(x)K^{0\rightarrow
2}_{w,w^{-1}}(X,Y)=K^{1\rightarrow1}_{w^{-1}}(Y,X^\ast). \label{Pimatrix}
\end{equation}
The simplest topology that contributes to these processes is that of
a cylinder, whose magnetic amplitude for $K^{1\rightarrow1}$ is
given in \eqref{cylamp}. The magnetic contribution of the same
cylinder to $K^{0\rightarrow2}$ is obtained by gluing a cap to a pair of pants.
Using the homotopy invariance, it is readily seen to be in
accordance with \eqref{Pimatrix}, the extra factor
$\Phi_{w,w^{-1}}(x)$ of the orientation reversing operation being
canceled by the inverse factor due to the interaction.
The commutation of the action of the quasi quantum group with
$K^{0\rightarrow 2}$ reads
\begin{equation}
\Delta(a)K^{0\rightarrow 2}=\epsilon(a)K^{0\rightarrow 2}.\label{commdelta}
\end{equation}
Reversing the orientation of the first string yields, using
Sweedler's notation for the coproduct (see appendix),
\begin{equation}
(\Pi\otimes\mbox{id})\left((a_{(1)}\otimes a_{(2)})K^{0\rightarrow2
}\right)=a_{(2)}K^{1\rightarrow 1}S\left(a_{(1)}\right).
\end{equation}
To derive the last equation, we have used \eqref{dualaction} that
states that the action of $a\in D_{\omega}[G]$ on an outgoing state
becomes, after orientation reversing, the dual action of $S(a)$ on
the corresponding incoming state. Then, \eqref{commdelta} translates
into
\begin{equation}
a_{(2)}S\left(a_{(1)}\right)=\epsilon(a).
\end{equation}
The latter follows from one of the defining relations of the
antipode
\begin{equation}
a_{(1)}\beta S(a_{(2)})=\epsilon(a)\beta,
\end{equation}
after acting with $S$ and using \eqref{squareofs} and its
antimorphism property $S(ab)=S(b)S(a)$, valid for arbitrary
quasi-Hopf algebras.

\subsection*{Associator on string states}
The quasi-quantum group  $D_{\omega}[G]$ is a quasi-triangular quasi
Hopf algebra. We refer to \cite{kassel} for some general background
on quasi Hopf algebras and their applications. We have collected in
the appendix a few results that are useful in our context. The
category of representations of such an algebra forms a quasi-tensor
(or braided monoidal) category  which is a far reaching
generalization of the category of representation of a group. Roughly
speaking, this is a category where tensor products are defined, with
an associativity law that holds only up to isomorphism, as well as
the exchange of the factors in the tensor products, which is
implemented using the braid group instead of the symmetric group.
This setting, termed quasi Hopf symmetry, has been proposed in
\cite{schomerus} as a natural framework that is versatile enough to
encompass all possible symmetries in physics. For example, a quasi
Hopf symmetry based on $D_{\omega}[G]$ appears in the study of
topological excitations coupled to non abelian Chern-Simons theory
in 2+1 dimensions \cite{bais}.

Let us now illustrate the quasi-associativity (associativity up to
isomorphims) of the tensor product in the context of interacting
strings.  As already mentioned, the tensor product of two
representations $\rho_{1}$ and $\rho_{2}$ of $D_{\omega}[G]$ has to
be defined using the coproduct rule \eqref{tens}. Consider now the
tensor product of three representations $\rho_{1}$, $\rho_{2}$ and
$\rho_{3}$. Because the coproduct is only quasi coassociative (see
\eqref{quasi666}), one has to distinguish between the two parenthesings
of the tensor products $\rho_{1}\otimes(\rho_{2}\otimes\rho_{3})$
and $(\rho_{1}\ot\rho_{2})\ot\rho_{3}$, these two representations
being intertwined by the image of $\Omega$.

\begin{figure}[h]\centering
\includegraphics[width=7cm]{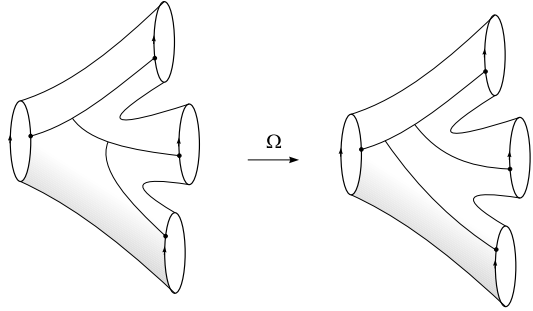}
\caption{Two different cuts related by the associator.}
\label{_assopant}
\end{figure}

\begin{figure}[t]\centering
\begin{tabular}{ccc}
\parbox{5cm}{\includegraphics[width=5cm]{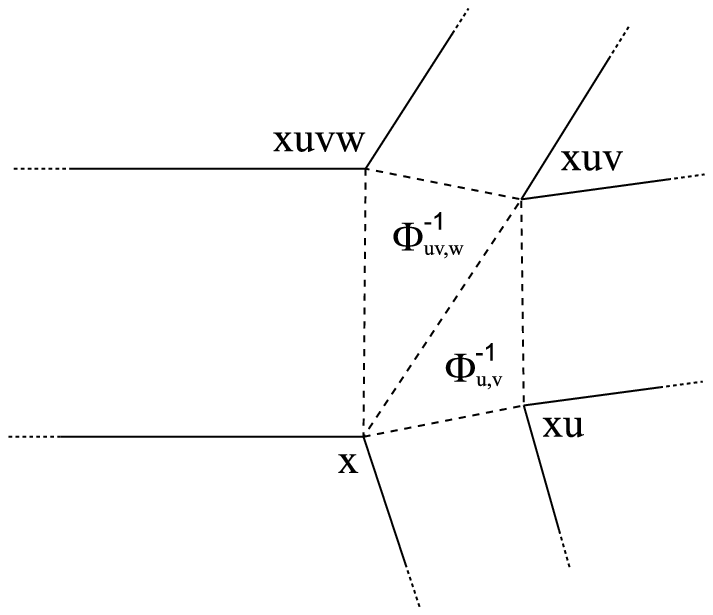}}
&$\leftrightarrow$&
\parbox{5cm}{\includegraphics[width=5cm]{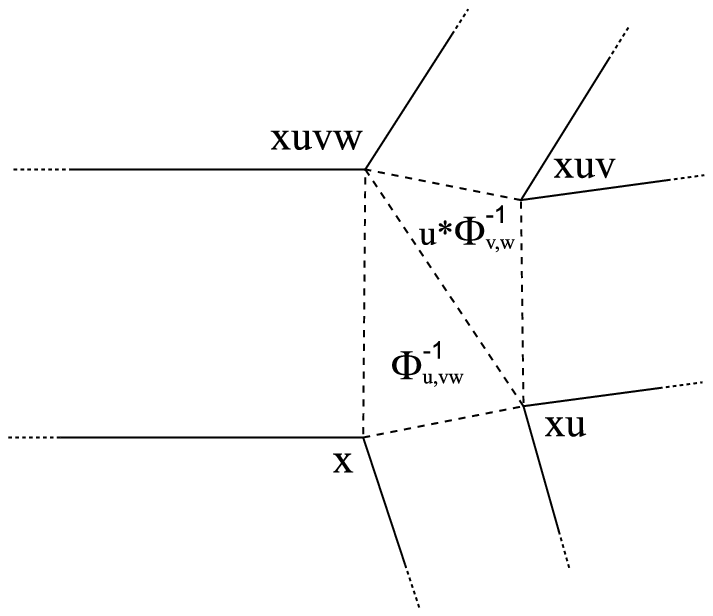}}
\end{tabular}
\caption{The fine structure of the branching point, with all
$\Phi^{-1}$'s
  evaluated at $x$.}
\label{_fine2}
\end{figure}

\begin{figure}[b]\centering
\begin{tabular}{ccc}
\parbox{3cm}{\includegraphics[width=3cm]{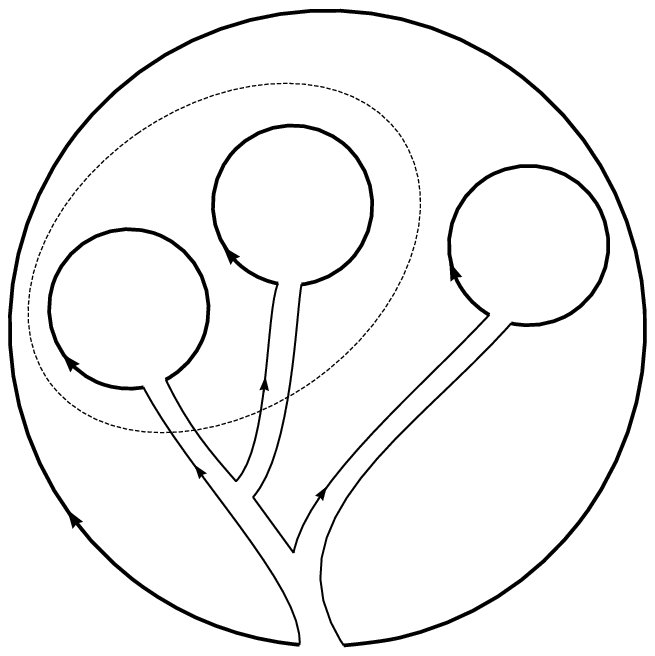}}&$\stackrel{\Omega}{\longrightarrow}$&
\parbox{3cm}{\includegraphics[width=3cm]{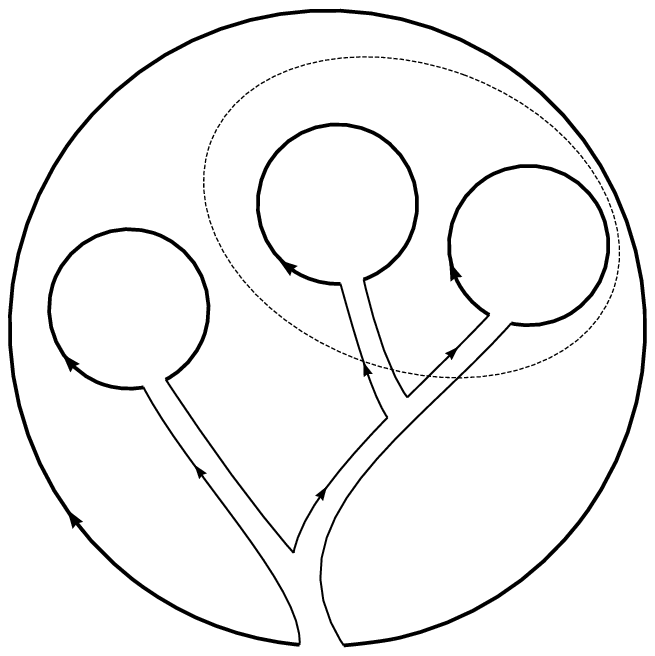}}
\end{tabular}
\caption{Quasi-associativity on the states.} \label{assos}
\end{figure}

The failure of associativity is very natural from the point of view
of the magnetic amplitudes. Consider the process in which one string
of winding $uvw$ decays into three of windings $u$, $v$ and $w$.
Such a process arises from the composition of two pairs of pants in
two different ways (see figure \ref{_assopant}).
%
To compare the two amplitudes, it is again useful to apply the
relation $\cD\Phi_{g,h}=g^{\ast}{\bf A}_{h}-{\bf A}_{gh}+{\bf
A}_{g}$ in such a way that the two branching points of the cut come
together. However, they cannot pass through each other and they
always remain in the same order.
%
We have depicted the fine structure of the branching point for the
$1\to 3$ decay in figure \ref{_fine2}.
%
The two amplitudes coincide except that the branching yields $\lb
\Phi_{uv,w}^{-1}\Phi_{u,v}^{-1}\rb_{i}$ for the first and $\lb
u^{\ast}\Phi_{v,w}^{-1}\Phi_{u,vw}^{-1}\rb_{i}$ for the second, with
$i$ being the index of the open set used to cover the branching
point. Thus, the second amplitude simply differs from the first one
by the phase factor $\omega_{u,v,w}$. This is in perfect agreement
with the quasi Hopf point of view, since the pattern of splittings
implies that the first process reads \be
\cH_{uvw}\rightarrow\cH_{uv}\ot\cH_{w}\rightarrow(\cH_{u}\ot\cH_{v})\ot\cH_{w},
\ee whereas the second is \be
\cH_{uvw}\rightarrow\cH_{u}\ot\cH_{vw}\rightarrow\cH_{u}\ot(\cH_{v}\ot\cH_{w}).
\ee The two processes end in two different tensor products,
differing only by their parenthesings. This statement is illustrated
in figure \ref{assos}, where we indicated the different
parenthesings on the Hilbert spaces by the dashed ellipses
encircling the corresponding states.
It is compatible with the action of magnetic translations, which
differ by the same phase for the two parenthesings.

The same result holds for more general amplitudes: the pattern of
interactions involved selects the parenthesing of the Hilbert spaces
appearing on the boundary. This is particularly clear if we consider
processes in which one string can decay into $n$ strings. The
branching of the cut yields a tree that encodes all the information
on the parenthesis to be used. The choice made for intermediate
states is irrelevant since changing their parenthesing always
produce two phases that cancel. Any change in the parenthesing can
be implemented by successive applications of the associator that
moves branches of the tree from one side of the corresponding
branching point to the other side. Using MacLane's coherence theorem
\cite{kassel}, the cocycle condition on $\omega$ implies that any
sequence of associators between two fixed parenthesing always yields
the same phase.

As already noticed for invariant states, the physical requirement of
the consistency of the orbifold forces $\omega$ to be trivial, so
that these complications do not arise in practice. In this case, the
quasi-quantum group reduces to the quantum double of the group $G$.
\medskip

\subsection{Tree level amplitudes}

\label{MCG}

Consider now all the tree level amplitudes describing the decay of one
string into $n$ others. From now on, we always
assume $\omega$ to be trivial, unless otherwise stated, so that there
is no need to keep track of the parenthesings. Any tree level  amplitude can
be obtained by gluing
cylinders and pants together. The resulting surface has a
mapping class group (i.e. those diffeomorphisms that are not connected to
the identity) generated by the Dehn twists of the cylinder and the
braidings of the pants. All these operators can be constructed using
the quasi-quantum group $D_{\omega}[G]$.

\subsection*{Dehn twist of the cylinder}
\begin{figure}[h]\centering
\includegraphics[width=9cm]{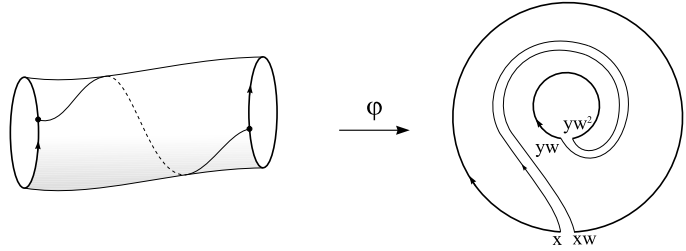}
\caption{The Dehn twist of the cylinder.}
\label{_dehn}
\end{figure}

\begin{figure}[h]\centering
\includegraphics[width=3cm]{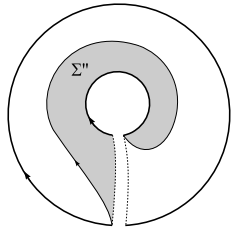}
\caption{The surface $\Sigma''$.}
\label{_thomot}
\end{figure}

The magnetic translations allow us to understand the transformation
of the cylinder amplitude under large diffeomorphisms. The
cylinder amplitude reads
\be
\cA_{JI}[\varphi]=\lb \mathrm{e}^{\ii\int_{\Sigma}{\bf B}+\ii\int_{x}^{y}{\bf A}_{w}}\rb_{JI}.
\ee
If we act with a Dehn twist on the cylinder, then the cut  $c$ is changed into
$c'$ which induces a corresponding change $\varphi\rightarrow\varphi'$.
In the target space ${\cal M}$, the transformation yields a new
worldsheet  $\Sigma'$ and the image of the new cut joins $x$ to
$y\!\cdot\!w$, as can be seen in figure \ref{_dehn}.
%
Therefore, we get an other amplitude $\cA_{JI}[\varphi']$ and the ratio of the
two amplitudes turns out to be
\be
\frac{\cA_{JI}[\varphi']}{\cA_{JI}[\varphi]}
=\lb \mathrm{e}^{\ii\int_{\Sigma''}\lp w^{\ast}{\bf B}-{\bf B}\rp
+\ii\int_{x}^{yw}{\bf A}_{w}-\ii\int_{x}^{y}{\bf A}_{w}}\rb_{JI}.
\ee
This follows from the fact that $\Sigma$ and $\Sigma'$ only differ by
the shaded surface $\Sigma''$ (see figure \ref{_thomot}),
whose contribution is lifted by $w$ in
$\cA_{JI}[\varphi']$ with respect to its contribution to
$\cA_{JI}[\varphi]$. After using $w^{\ast}{\bf B}={\bf B}+\cD {\bf A}_{w}$ in the explicit
expression of the ratio, only boundary terms remain. Two of them
cancel with the integrals along the cut and we are left with
\be
\frac{\cA_{JI}[\varphi']}{\cA_{JI}[\varphi]}
=\lb \mathrm{e}^{\ii\int_{y}^{yw}{\bf A}_{w}}\rb_{J}.
\ee
Recall the translation by $w$ of a string $Y$ with winding $w$,
\be
\lb T_{w}^{w}\Psi\rb_{J}(Y)=
\lb \mathrm{e}^{-\ii\int_{y}^{yw}{\bf A}_{w}}\rb_{J}\Psi_{J}(Y).
\ee
Thus, the action of the Dehn twist simply amounts to translating the
outgoing string by its own winding. Starting with the twisted cylinder
one constructs the twisted propagator $K'$ and the comparison of the
magnetic amplitudes for $\varphi$ and $\varphi'$ shows that
$T_{w}^{w}K_{w}^{'}=K_{w}$. Because of the commutation of $T_{w}^{w}$
and $K_{w}$, the twist can as well be implemented on the incoming string.
%

\subsection*{Braiding of the pair of pants}
\begin{figure}[h]\centering
\includegraphics[width=8cm]{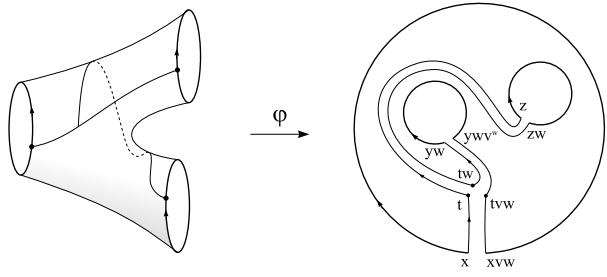}
\caption{The braiding.}
\label{_tresse}
\end{figure}

\begin{figure}[b]\centering
\includegraphics[width=7cm]{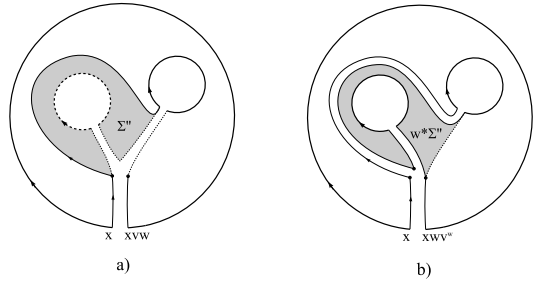}
\caption{The surface $\Sigma''$ and its lift by $w$.}
\label{_tresse_homot}
\end{figure}

Consider now the large diffeomorphism of the pant that induces a
braiding of the cut as in figure \ref{_tresse}.
%
%
The braiding induces the change $\varphi\rightarrow\varphi'$ associated
to the new amplitude
\be
\cA_{KJ,I}[\varphi']=\lb \mathrm{e}^{\ii\int_{\Sigma'}{\bf B}
+\ii\int_{x}^{t}{\bf A}_{vw}
+\ii\int_{tw}^{yw}{\bf A}_{v^w}
+\ii\int_{t}^{z}{\bf A}_{w}}
\;\Phi^{-1}_{w,v^{w}}(t)\rb_{KJ,I}
\ee
where the line integrals involve the cuts in figure \ref{_tresse}. The
new amplitude $\cA_{KJ,I}[\varphi']$ differs from the old one
$\cA_{JK,I}[\varphi]$ given in \eqref{pant} by the ordering of the
windings ($(v,w)\rightarrow (w,v^{w})$) and by the shaded surface
(see figure \ref{_tresse_homot}) which
is removed from $\Sigma$, lifted by $w$ and reinserted into $\Sigma$ to
form $\Sigma'$.
%

Thus, the ratio of the two amplitudes is
\be
\frac{\cA_{KJ,I}[\varphi']}{\cA_{JK,I}[\varphi]}=
\lb\Gamma_{v,w}^{-1}(y)\;\mathrm{e}^{\ii\int_{y}^{yv}{\bf A}_{w}}\rb_{J}\label{ratiobraid}.
\ee
This matches exactly the phase that appears in the operator
$T_{w}^{v}$. Therefore, the braiding of the pants induces a change of
the magnetic amplitude $\cA_{JK,I}[\varphi]\rightarrow \cA_{KJ,I}[\varphi']$
that corresponds to the braid group action derived from the general
theory of quasi-triangular quasi Hopf algebras (see appendix). Indeed,
the quasi-quantum group $D_{\omega}[G]$ is equipped with an
$\cR$-matrix defined by
\be
\cR=\mathop{\sum}\limits_{v,w}
T_{e}^{v}\ot T_{v}^{w}.
\label{Rmatrix}
\ee
When acting on a tensor product of two states, it simply leaves the
first string invariant and translates the second by the winding of the
first. The corresponding action of the braid group with two strands $B_{2}$
is defined by its generator $\sigma=\tau\circ{\cal R}$, where $\tau$
is the flip.
If we restrict ourselves to strings with fixed windings, this
induces a map $\sigma_{v,w}:\, {\cal H}_{w}\otimes{\cal
  H}_{v^{w}}\rightarrow{\cal H}_{v}\otimes{\cal H}_{w}$ defined by
$\sigma_{v,w}=\tau\circ(1\otimes T_{w}^{v})$. Then, the same reasoning
as for the twist of the cylinder applies: The braiding yields another
pant ${\cal S'}$ whose field $\varphi'$ defines a propagator $K'$
related to $K$ by $\sigma_{v,w}K'_{w,v^{w}}=K_{v,w}$.

Note that thanks to the quasi-triangularity condition
\be
\tau\circ\Delta=\cR\,\Delta\cR^{-1},
\ee
the braid group action commutes with the stringy magnetic translations,
provided the latter act on tensor products using the coproduct rule.
\medskip

When more than two factors are involved
(recall that, for the moment, we take $\omega$ to be trivial, i.e.
the Drinfel'd associator is also trivial),
the exchange of the factors
is governed by the braid group with $N$ strands $B_{N}$. The latter is
generated by the operators $\sigma_{i}=\tau\circ\cR$ that exchange the two adjacent factors located at the $i^{\mathrm{th}}$ and
$(i+1)^{\mathrm{th}}$ place in the tensor product. The defining
relations of $B_{N}$, i.e.
\be
\left\{
\begin{array}{rclr}
\sigma_{i}\sigma_{j}&=&\sigma_{j}\sigma_{i}&\mbox{if}\quad|i-j|>1\cr
\sigma_{i}\sigma_{i+1}\sigma_{i}&=&\sigma_{i+1}\sigma_{i}\sigma_{i+1}&
\mbox{otherwise},
\end{array}
\right.
\label{braidrelation}
\ee
hold thanks to the Yang-Baxter equation (see appendix),
\begin{equation}
{\cal R}_{12}{\cal R}_{13}{\cal R}_{23}=
{\cal R}_{23}{\cal R}_{13}{\cal R}_{12}.\label{YB}
\end{equation}
Together with the Dehn twists of the cylinder, the braid group
$B_{N}$ generates the mapping class group of the pants with one
incoming and $N$ outgoing strings.

\subsection*{Braided operad}
The tree level amplitudes corresponding to the decay
of a single string into $N$ others are conveniently
visualized as follows.
\begin{itemize}
\item
Draw a large cut circle that stands for the incoming string.
\item
Draw $N$ smaller circles inside the larger one that represent the
outgoing strings.
\item
Relate the endpoints of the incoming string to those of the outgoing
ones by a tree, allowing for twists around the circles and braids
around two neighbouring circles.
\end{itemize}

\begin{figure}[h]\centering
\includegraphics[width=6cm]{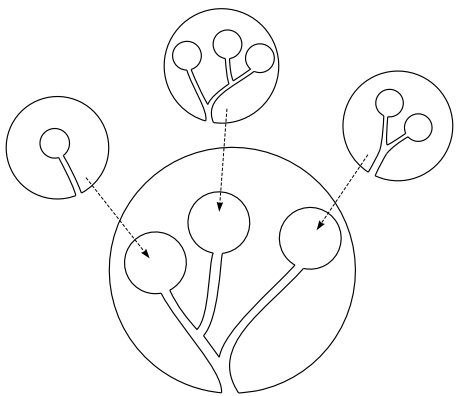}
\caption{The braided operad structure of tree level amplitudes.}
\label{opstr}
\end{figure}

An amplitude for a process $1\rightarrow N$ can be glued with $N$
amplitudes for $1\rightarrow n_{N}$ so that the result is an amplitude
pertaining to the process $1\rightarrow\sum_{i=1}^{N}n_{i}$. In figure
\ref{opstr}, we give a simple example of gluing three amplitudes into
the corresponding slots of a first one.
More generally, denoting by $\Delta_{n}$ the set of amplitudes with $n$ outgoing
strings, the gluings define composition laws
\be
\gamma_{N}:\,\Delta_{N}\times\Delta_{n_{1}}\times\cdots\times\Delta_{n_{N}}\rightarrow\Delta_{n_{1}+\cdots+n_{N}}.
\ee
The sets $\Delta_{n}$ carry an obvious representation of the braid
group $B_{n}$, with generators $\sigma_{i}$ acting in the same manner
as for the pair of pants. The action is compatible with the
gluing, in the sense that braiding before gluing is equivalent to
braiding after gluing. Altogether, this means that the decaying
amplitudes form a braided operad \cite{braided}.

In all the preceeding discussion we have assumed that the 3-cocycle
$\omega$ is trivial. If this is not the case, the braid group action
still makes sense, but some care is required because of the lack of
associativity of the tensor product. For the action of $B_{N}$ defined
using its generators $\sigma_{i}$ one has to make a repeated use of the
associator to make sure that the parenthesings do not seperate the two
factors being exchanged. Therefore, when composing $\sigma_{i}$ and
$\sigma_{j}$, one has to insert suitable representations of the
associator (see appendix). The defining relations of the braid group
\eqref{braidrelation} then follow from the quasi Yang-Baxter
equation \eqref{qybe}, which is a modification of the Yang-Baxter equation
\eqref{YB} in order to take into account the Drinfel'd associator.

\subsection{Loop amplitudes}
\begin{figure}[h]\centering
\includegraphics[width=6.5cm]{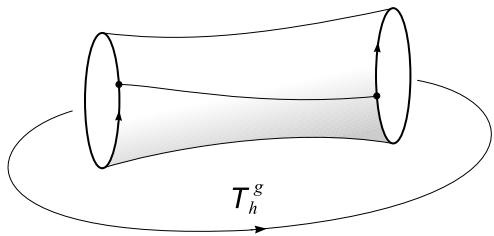}
\caption{Obtaining a torus amplitude from a cylinder.}
\label{toram}
\end{figure}

The simplest loop amplitude to consider is the torus,
that can be obtained from the cylinder by gluing the two boundaries
using a stringy magnetic translation as depicted in figure \ref{toram}.
The amplitude reads
\be
\cA[\varphi]=\lb \mathrm{e}^{\ii\int_{\Sigma}{\bf B}+\int_{x}^{x\cdot
    h}{\bf A}_{g}-\ii\int_{x}^{x\cdot g}{\bf A}_{h}}\,\Gamma_{g,h}(x)\rb,
\ee
where $g$ and $h$ are two mutually commuting elements of $G$.  It is
obtained from the cylinder amplitude of a propagating string of
winding $g$ by adding the extra phase corresponding to the translation
by $h$ of a string of winding $g$, and should be though of as being
inserted into a functional integral contributing to the trace of
$(T_{h}^{g})K_{g}$, which is further summed over $g$ and $h$ to
provide the torus partition function.

Because there is no boundary, it does not depend at all on the
triangulation used in its computation. Nevertheless, we write it into brackets
to remember that its actual definition relies on some
triangulation. Besides, it is invariant under
homotopic changes of the cut, diffeomorphisms connected to the
identity and gauge transformations. These properties  are valid whether
$\omega$ is trivial or not.

However, when $\omega$ is non trivial, the torus amplitude is plagued
by  global anomalies: it receives extra phases under global translations
and large diffeomorphisms of the torus. Under a global translation
$\varphi\rightarrow\varphi\!\cdot\!k$, the amplitude changes according to
\be
\cA[\varphi\!\cdot\! k]=
\frac{\omega_{k,h^{k},g^{k}}\,\omega_{g,k,h^{k}}\,\omega_{h,g,k}}
{\omega_{k,g^{k},h^{k}}\,\omega_{g,h,k}\,\omega_{h,k,g^{k}}}\,\cA[\varphi].
\ee
Using the relation \eqref{product} and its geometrical intepretation
given in figure \ref{_prism1},  this combination of 3-cocycles can be
understood as a way of chopping the parallelepiped build out of $g$,
$h$ and $k$ into six tetrathedra. When all three group elements
commute, this matches (up to a global sign) the topological action for
a 3-torus in finite group topological field theory \cite{dijkgraafwitten}.

The group of large diffeomorphisms of the torus is generated by the
modular transformations $T$ and $S$ that induce the following changes
in the group elements along the cut,
\be
\begin{array}{ccc}
T:\quad
\la
\begin{array}{lcl}
g&\rightarrow&gh\\
h&\rightarrow& h
\end{array}
\right.
&&
S:\quad
\la
\begin{array}{lcl}
g&\rightarrow&h^{-1}\\
h&\rightarrow&g
\end{array}
\right.
\end{array}.
\ee
Again, the amplitude changes by extra phases depending on the
3-cocycle $\omega$. Indeed, $T$ acts as
\be
\cA[\varphi]\rightarrow\cA[T\varphi]=\omega_{h,g,h}\,\cA[\varphi]
\ee
and $S$ as
\be
\cA[\varphi]\rightarrow\cA[S\varphi]=
\frac{\omega_{h,g,h^{-1}}}
{\omega_{h,h^{-1},g}\,\omega_{g,h,h^{-1}}}
\,\cA[\varphi].
\ee
This behaviour of the torus amplitude and of all the previous tree
level amplitudes under global translations and large diffeomorphisms
yields relations that  are very similar to the ones
in \cite{bantay}, derived in the context of CFT. In fact, this last
paper deals with algebraic properties of the conformal blocks in an
orbifold theory with group $G$. It is rather amazing that the formulae
pertaining to the transformation of conformal blocks are so close to
the ones derived here for magnetic amplitudes, especially because our
derivation relies solely on the geometry of the Kalb-Ramond field and
does not refer to CFT.
\medskip

More generally, we expect all loop amplitudes to be flawed by these
global anomalies under translation and global diffeomorphisms. This is
due to the insertion of the gauge fields ${\bf A}_{g}$ and $\Phi_{g,h}$
along the internal cuts, which is required by gauge invariance. Any
global change induces a change of these gauge fields which is only
partially compensated along the two boundaries of the cut. This is
another clue that the orbifold theory is consistent only for trivial $\omega$.

If we assume that $\omega$ is trivial, then one can construct the
magnetic amplitude for an arbitrary surface as follows. At the tree
level, we construct the amplitude by gluing together cylinders and
pairs of pants as given in \eqref{fullamplitude} and
\eqref{pant}. For the latter it is crucial to insert a factor of
$\Phi^{-1}_{v,w}$ when a string of winding $vw$ splits into two strings
of windings $v$ and $w$ and $\Phi_{v,w}$ when they join. Loop
amplitudes are obtained by
further gluing together some incoming and outgoing strings using the
operators $T_{g}^{w}$. At the level of magnetic amplitudes, this
translates into insertions of the phases $\Upsilon_{g}^{w}$ when a
string of winding $w$ is glued with its lift by $g$. Then, the
triviality of $\omega$ ensures that the amplitude is independent of the
patterns of joining and splitting, as well as under global
translations and large diffeomorphisms.

\subsection{Relation to discrete torsion}

\label{discretetorsion}

Let us now assume that $\omega$ is trivial, so that the orbifold
exists, and come to grips with discrete torsion \cite{vafa}. In string
theory, discrete torsion refers to a set of group dependent phases
that weight the orbifold amplitudes. Its geometrical nature has been
unveiled in \cite{sharpe}, where it is shown that it corresponds to the
ambiguity appearing in the lift of the orbifold group action to the
fields. Let us show how this ambiguity affects the definition of the
stringy magnetic translations.

Recall that there is a constant ambiguity in the definition of $\Phi$
given in \eqref{ambiguityalpha}. Indeed,
$\Phi$ is defined up to multiplication by a constant group cochain
$\alpha$ and can be replaced by $\Phi'=\Phi\,\alpha$, leaving the
fields ${\bf A}$ and ${\bf B}$ untouched. This induces
the change $\omega\rightarrow\omega'=\omega\delta\alpha$, so that $\omega$ is
preserved if and only if $\alpha$ is a group cocycle. If we display
the group elements,
\begin{equation}
\Phi_{g,h}\rightarrow\Phi'_{g,h}=\Phi_{g,h}\alpha_{g,h}
\end{equation}
where $\alpha$ fulfills the 2-cocycle condition
\begin{equation}
\alpha_{h,k}\alpha_{g,hk}=\alpha_{gh,k}\alpha_{g,h}.
\label{cocyclealpha}
\end{equation}

Because the stringy magnetic translations $T_{g}^{w}$ defined in
\eqref{definition} depend on $\Phi$ through $\Gamma$ (see definition
\eqref{defGamma}), the 2-cocycle $\alpha$ induces the change \be
T_{g}^{w}\rightarrow (T')_{g}^{w}=\epsilon_{w,g}\, T_{g}^{w}, \ee
with \be
\epsilon_{w,g}=\frac{\alpha_{g,w^{g}}}{\alpha_{w,g}}.\label{discrete}
\ee When $\alpha=\delta\beta$ is a coboundary, this simply amounts
to multiply wave functions in ${\cal H}_{w}$ by the global phase
$\beta_{w}$, so that the ambiguity is parametrized by the cohomology
class of $\alpha$. Note that the expression of dicrete torsion can
also be interpreted as a transgression of the 2-cocycle $\alpha$
(see \eqref{exampletrans}.)

The operators $T_{g}^{w}$ and $(T')_{g}^{w}$ obey the same
multiplication law given in \eqref{product}, since their product only
involves $\omega$ which is left unchanged by a 2-cocycle. Replacing
$(T')_{g}^{w}$ in terms of $T_{g}^{w}$ and $\epsilon_{w,g}$ in the
product law \eqref{product} leads to the
identity
\begin{equation}
\epsilon_{w,g}\epsilon_{w^{g},h}=\epsilon_{w,gh},
\end{equation}
that can be checked directly by repeated use of the cocycle condition
\eqref{cocyclealpha}. Thus, at fixed $w$ the application
$g\mapsto\epsilon_{w,g}$ defines a 1-cocycle on the normalizer $N_{w}$
of $w$. As we shall see below, $\epsilon_{w,g}$ naturally identifies with
discrete torsion.

Nevertheless, the coproduct is not similar in form for $(T')_{g}^{w}$
and $T_{g}^{w}$. Indeed,
\begin{equation}
\Delta({(T')}_{g}^{u})=\mathop{\sum}\limits_{vw=u}\,
\frac{\alpha_{v^{g},w^{g}}}{\alpha_{v,w}}\,
\frac{\omega_{v,w,g}\,\omega_{g,v^{g},w^{g}}}{\omega_{v,g,w^{g}}}\,
{(T')}_{g}^{v}\ot {(T')}_{g}^{w},
\end{equation}
as should have been expected since the definition of the coproduct depends on
the interaction which involves $\Phi$ as required by invariance
under  secondary gauge transformation (see the discussion at the
begining of section \ref{interaction}). To recover an equation similar
to \eqref{coproduct}, one has to trade $\Delta$ for a new coproduct,
\begin{equation}
\Delta\rightarrow\Delta'={\cal F}\Delta{\cal F}^{-1},
\end{equation}
with
\begin{equation}
{\cal F}=\mathop{\sum}\limits_{u,v\in G}\alpha^{-1}_{u,v}\,
T_{e}^{u}\otimes T_{e}^{v}.
\end{equation}
This operation corresponds to a Drinfel'd twist of the quasi-quantum
group $D_{\omega}[G]$ (See appendix). It further induces a change of
the ${\cal R}$ matrix involved in the braiding,
\begin{equation}
{\cal R}\rightarrow{\cal R}'
=\mathcal{F}_{21}\mathcal{R}\mathcal{F}^{-1}_{12}
=\mathop{\sum}\limits_{v,w\in G}
\epsilon_{w,v}\;T_{e}^{v}\otimes T_{v}^{w},
\end{equation}
but leaves the associator $\Omega$ invariant because of the cocycle
condition \eqref{cocyclealpha}.

Accordingly, discrete torsion is an
ambiguity on the lift of the orbifold group action to the
fields. This induces an ambiguity on the operators $T_{g}^{w}$ that
lift the group action to the twisted states. Then, any two choices
$T_{g}^{w}$ and ${(T')}_{g}^{w}$ of such operators  generate
quasi-quantum groups related by a Drinfel'd twist.  They are equivalent
as far as their representations are considered \cite{kassel}. This is
similar to the discussion pertaining to vacuum angles in section
\ref{particleproj}, where the ambiguity on the lift of the group
action to particle states were classified by a one cocycle.

If we restrict ourselves to group elements $g$, $h$ that commute with
$w$, then $\epsilon_{w,g}$ fulfills
\be
\la
\begin{array}{rcl}
\epsilon_{w,gh}&=&\epsilon_{w,g}\epsilon_{w,h}\, ,\cr
\epsilon_{g,w}&=&\lp\epsilon_{w,g}\rp^{-1},\cr
\epsilon_{g,g}&=&1.
\end{array}
\right.
\label{conditionsondc}
\ee
These are exactly the conditions imposed on discrete torsion. The
relation between discrete torsion and
group cohomology goes back to the very first paper on the subject
\cite{vafa}, but we find it interesting to show how it appears in the
present context.

If we further restrict ourselves to an abelian group, then the
projector defined by
\begin{equation}
P_{w}=\mathop{\sum}\limits_{g\in G}\epsilon_{w,g}\,T_{g}^{w},
\end{equation}
projects the twisted sector labeled by $w$
onto states that are invariant under the action of $G$ up to a
phase.  Indeed,
\begin{equation}
T_{g}^{w}P_{w}=(\epsilon_{w,g})^{-1}P_{w},
\end{equation}
so that the states of the orbifold string theory with discrete torsion
belong to $P_{w}{\cal H}_{w}$.

Moreover, the
braid group action (with the ${\cal R}$-matrix defined in
\eqref{Rmatrix}) on the tensor product of two such states fulfills
\begin{equation}
\sigma\left( P_{w}\otimes P_{v}\right)=
(\epsilon_{v,w})^{-1}\left( P_{v}\otimes P_{w}\right).
\end{equation}
Consequently, if we act twice with the braiding,
\begin{equation}
\sigma^{2}\left( P_{v}\otimes P_{w}\right)=
(\epsilon_{v,w}\epsilon_{w,v})^{-1}\left( P_{v}\otimes  P_{w}\right)=
P_{v}\otimes  P_{w},
\end{equation}
because of the second equation in \eqref{conditionsondc}.
Therefore the braid group action reduces to the standard symmetric
group action after projection, in accordance with the bosonic nature
of the orbifold theory. Similar results are expected to hold for a non
abelian group, with twisted sectors labeled by conjugacy classes of
the windings.

\section{Conclusion}

With a view to the construction of an orbifold string theory ${\cal M}/G$
in the presence of a $G$-invariant 3-form $H$ on ${\cal M}$  we have
studied the operators $T_{g}^{w}$ that realize the action of the group
$G$ on the twisted sectors.

The results presented here can be summarized into two main points.

The de Rham, {\Cech} and group coboundaries are combined into a
tricomplex that provides a convenient framework to deal with the
invariance of locally defined higher  gauge fields, as the $B$-field of a
string. Starting with an invariant 3-form
$H$, this provides a sequence of \Cech-de Rham gauge fields of decreasing
degree, ending in a constant group 3-cocycle $\omega$.  Though we confined
ourselves to the simplest case of a bosonic string theory, the same
techniques could be applied to a systematic study of the higher rank
gauge fields appearing in supersymmetric theories. We have illustrated
this in the case of the M-theory 3-form potential $C$.

Using the previously defined gauge fields, the
operators $T_{g}^{w}$ are constructed by requiring that they commute
with the propagation of a single string. Besides, they generate the
quasi-quantum group $D_{\omega}[G]$, introduced by R. Dijkgraaf,
V. Pasquier and P. Roche \cite{dijkgraaf}, whose coproduct is derived from
the tree level interactions of the twisted sectors. However, the existence of
invariant states and the definition of the loop amplitudes
requires the 3-cocycle $\omega$ to be trivial in order to construct an
orbifold. This confirms, from another viewpoint, the results presented
in \cite{sharpe} and the geometric analysis performed in \cite{gawedzki}.
In this case, a general procedure for the construction of the magnetic
amplitudes is outlined: Starting with a tree level amplitude one
identifies some of the boundaries using the operators $T_{g}^{w}$.
Nevertheless, let us stress that even if $\omega=1$ the phase
factor $\Upsilon_{g}^{w}(X)$ has to be inserted in the transformtion law
of the wave function of the string and cannot, in general, be gauged
away. Of course, if this phase is to be used in a realistic model, one
has to change the basis and express $\Upsilon^{w}_{g}$ as an operator made
out of the string's oscillators.

As a general conclusion it can be said that the quasi-quantum group
$D_{\omega}[G]$ provides a stringy generalization of the projective
group representation associated with the dynamics of a particle in an
invariant $B$-field, based on a 3-cocycle instead of a 2-cocycle.
As such, we expect $D_{\omega}[G]$ (or some of
its generalizations) to play a significant role as symmetries of
extended objects, just as projective group representations appear when
lifting the orbifold group action to the non abelian gauge fields on coinciding
D-branes, in the presence of discrete torsion \cite{douglas}. This could
occur in the context of M-theory, when
lifting the group action to the non abelian gauge fields introduced in
\cite{jurco}, in the presence of discrete torsion identified as a 3-cocycle
in \cite{sharpeMtheory}.



\newcommand{\alg}{\mathcal{B}}  
\newcommand{\cat}{\mathcal{C}}  

\appendix
\section{Quasi Hopf algebras}
\subsection*{General construction}
The algebraic structure we encountered is the structure of a quasi Hopf algebra.
Generally speaking, a quasi Hopf algebra is a Hopf algebra $\alg$
where some defining relations are relaxed. The most important difference is
the existence of the {\it Drinfel'd associator}
$\Omega\in\alg\otimes\alg\otimes\alg$. One can then relax
coassociativity to quasi-coassociativity which has a quasi-bialgebra structure
as a consequence. Quasi-coassociativity reads
\begin{equation} (\mathrm{id}\otimes\Delta)\circ\Delta(b)=\Omega\left[
(\Delta\otimes\mathrm{id})\circ\Delta(b)\right]\Omega^{-1}
\qquad\forall b\in\alg.
\end{equation}
Additionally, the Drinfel'd associator fulfills an equation related to the
pentagon axiom of the associated tensor category of modules over the algebra,
namely
\begin{equation} (\mathrm{id}\otimes\mathrm{id}\otimes\Delta)(\Omega)(\Delta\otimes\mathrm{id}\otimes\mathrm{id})(\Omega)=\Omega_{234}
 \left[(\mathrm{id}\otimes\Delta\otimes\mathrm{id})(\Omega)\right]\Omega_{123},
\end{equation}
where we used the convention that for an element $b=b_1\otimes b_2\otimes b_3\in\alg^{\otimes 3}$,
the notation $b_{ijk}$ means to embed the element $b$ into $\alg^{\otimes n}$
by repeated insertion of $1_\alg$ and possibly reshuffling the components $b_i$. For
example
\begin{equation}
b_{ijk}:=1_\alg \otimes...\otimes b_1 \otimes 1_\alg \otimes...\otimes b_3 \otimes...\otimes b_2 \otimes...\otimes 1_\alg \quad
\in\alg^{\otimes n}
\end{equation}
is the element, where $b_1$ is in the $i^{\mathrm{th}}$ position, $b_2$ in the
$j^{\mathrm{th}}$, $b_3$ in the $k^{\mathrm{th}}$
and the rest is $1_\alg$. Here, $b\in\alg^{\otimes 3}$ is of course just an example, and the same
notation is used for general elements $\alg^{\otimes k}$ which are embedded into
$\alg^{\otimes n}$ for $k\le n$.
\medskip

Apart from that, the counit constraints for the coalgebra structure can also be relaxed, by
$(\epsilon\otimes\mathrm{id})\circ\Delta(b)=l^{-1}b\,l$, $(\mathrm{id}\otimes\epsilon)\circ\Delta(b)=r^{-1}b\,r$ and
$(\mathrm{id}\otimes\epsilon\otimes\mathrm{id})(\Omega)=r\otimes l^{-1}$ with two elements
$r,l\in\alg$, but we will not deal with that in the sequel, since in our case, $r=l=1_\alg$.
\medskip

For the antipode $S$, the standard relation is modified by the requirement of the existence
of two elements $\alpha,\beta\in\alg$, such that
\begin{equation}
\begin{array}{ll} i)& S(b_{(1)})\,\alpha\, b_{(2)}=\epsilon(b)\alpha \\
ii)& b_{(1)}\,\beta\, S(b_{(2)})=\epsilon(b)\beta \end{array}
\qquad \forall b\in \alg\label{antipode}
\end{equation}
and
\begin{equation}
\sum \varphi_1\,\beta\, S(\varphi_2)\,\alpha\,\varphi_3=
\sum S(\bar{\varphi}_1)\,\alpha\,\bar{\varphi}_2\,\beta\, S(\bar{\varphi}_3)=1_\alg,
\end{equation}
with the notation $\Omega=\sum \varphi_1\otimes\varphi_2\otimes\varphi_3$ and
$\Omega^{-1}=\sum \bar{\varphi}_1\otimes\bar{\varphi}_2\otimes\bar{\varphi}_3$. In (\ref{antipode}),
recall Sweedler's notation, i.e. $\Delta(b)=b_{(1)}\otimes b_{(2)}$.
\medskip

In standard parlance, a bialgebra is quasi-cocommutative, if there is an invertible element
$\mathcal{R}\in\alg\otimes\alg$, called $\mathcal{R}$-matrix, such that for all $b\in\alg$,
$\Delta^{\mathrm{op}}(b)=\mathcal{R}\Delta(b)\mathcal{R}^{-1}$. It is then called braided, if
furthermore the braid relations  $(\mathrm{id}\otimes\Delta)\mathcal{R}=\mathcal{R}_{13}\mathcal{R}_{12}$ and
$(\Delta\otimes\mathrm{id})\mathcal{R}=\mathcal{R}_{13}\mathcal{R}_{23}$ are satisfied which have
the Yang-Baxter equation (YBE)
$\mathcal{R}_{12}\mathcal{R}_{13}\mathcal{R}_{23}=\mathcal{R}_{23}\mathcal{R}_{13}
\mathcal{R}_{12}$ as a consequence.
A \underline{quasi}-bialgebra can also be braided:
The definition involves the braid relations and the Yang-Baxter equation.
Whereas the quasi-cocommutativity stays untouched, i.e.
$\Delta^{\mathrm{op}}(b)=\mathcal{R}\Delta(b)\mathcal{R}^{-1}$, the other ones will
mix up with the Drinfel'd associator $\Omega$, i.e.
\begin{equation}
\begin{array}{ll}
  i)&(\mathrm{id}\otimes\Delta)\mathcal{R}=\Omega^{-1}_{231}\mathcal{R}_{13}\Omega_{213}\mathcal{R}_{12}\Omega^{-1}_{123}\\
  ii)&(\Delta\otimes\mathrm{id})\mathcal{R}=\Omega_{312}\mathcal{R}_{13}\Omega^{-1}_{132}\mathcal{R}_{23}\Omega_{123}\end{array}
\end{equation}
and
\begin{equation}
\mathcal{R}_{12}\Omega_{312}\mathcal{R}_{13}\Omega^{-1}_{132}\mathcal{R}_{23}\Omega=
\Omega_{321}\mathcal{R}_{23}\Omega^{-1}_{231}\mathcal{R}_{13}\Omega_{213}\mathcal{R}_{12}.
\label{qybe}
\end{equation}
The YBE is now called {\it Quasi-}Yang-Baxter equation (QYBE). A quasi Hopf algebra
is braided, if the underlying quasi bialgebra is.
\medskip

\subsection*{Drinfel'd twist}
Generally, a Drinfel'd twist is an equivalence relation on quasi Hopf algebras such that
the category of modules of two equivalent quasi Hopf algebras are tensor equivalent. In
the literature, it is also called a {\it gauge transformation} of the Hopf algebra.
The twist is constructed with the aid of an invertible element $\mathcal{F}\in\alg\otimes\alg$
such that $(\epsilon\otimes\mathrm{id})\mathcal{F}=(\mathrm{id}\otimes\epsilon)\mathcal{F}=1$.
One can then define a new quasi Hopf algebra, whose quasi bialgebra structure is given by
the twisted coproduct as well as a new Drinfel'd associator:
\begin{equation}
\left\{ \begin{array}{l}
\Delta_{\mathcal{F}}:\alg\to\alg\otimes\alg,\quad \alg\ni b\mapsto\Delta_\mathcal{F}(b)=
\mathcal{F}\Delta(b)\mathcal{F}^{-1}\\
\Omega_{\mathcal{F}}=\mathcal{F}_{23}(\mathrm{id}\otimes\Delta)[\mathcal{F}]
\Omega(\Delta\otimes\mathrm{id})[\mathcal{F}^{-1}]\mathcal{F}^{-1}_{12}
\end{array}\right.,
\end{equation}
whereas the counit stays the same. It is then the triple
$(\epsilon,\Delta_{\mathcal{F}},\Omega_{\mathcal{F}})$ which defines the gauge transformed
quasi bialgebra structure on $\alg$, now denoted by $\alg_{\mathcal{F}}$.

Recall that a quasi Hopf structure comes with elements $\alpha,\beta\in\alg$ as a modification
of the standard antipode relations. If $\alpha$ and $\beta$ are the corresponding elements
in $\alg$, then a Drinfel'd twist carries them over to $\alg_\mathcal{F}$. With the notation
$\mathcal{F}=\sum f_1\otimes f_2$ and $\mathcal{F}^{-1}=\sum \overline{f_1}\otimes\overline{f_2}$,
define new versions $\alpha_\mathcal{F}$ and $\beta_{\mathcal{F}}$ by
\begin{equation}
\left\{ \begin{array}{l}
\alpha_\mathcal{F}=\sum S(\overline{f_1})\,\alpha\,\overline{f_2}\\
\beta_\mathcal{F}=\sum f_1\,\beta\,S(f_2)
\end{array}\right..
\end{equation}
The quasi Hopf structure of the new collection $\alg_\mathcal{F}$, $\epsilon$, $\Delta_\mathcal{F}$,
$\Omega_\mathcal{F}$, $\alpha_\mathcal{F}$ and $\beta_\mathcal{F}$ is then complete.
\medskip

If $\alg$ additionally comes with a braiding $\mathcal{R}$,
then $\alg_\mathcal{F}$ is also equipped with
a quasi-triangular structure,
\begin{equation}
\mathcal{R}_{\mathcal{F}}=\mathcal{F}_{21}\mathcal{R}\mathcal{F}^{-1}_{12}
\end{equation}
fulfilling all necessary relations of the $\mathcal{R}$-matrix and braid relations using
the transformed versions $\Delta_{\mathcal{F}}$ and $\Omega_{\mathcal{F}}$.

\subsection*{Modules as tensor categories}
In the sequel, we focus on
the aspect that the category of modules over a bialgebra is a tensor category $\cat$,
that is, it comes naturally equipped with a monoidal structure, which is a functor
$\otimes:\cat\times\cat\to\cat$. In what follows, $\otimes$ is called
tensor product and the category comes with an associativity constraint, i.e. for all
objects $U,V,W\in\cat$, there is a
functorial isomorphism $\Phi_{U,V,W}:(U\otimes V)\otimes W\to U\otimes (V\otimes W)$
compatible with the left (repectively right) unity constraint of $\cat$ and fulfilling the
pentagon axiom. More precisely,
suppose $M_1, M_2$ are modules over an algebra $\alg$. Then, there is
a $\alg\otimes\alg$-module structure on the tensor product $M_1\otimes M_2$, given canonically by
$(a\otimes b)(m_1\otimes m_2)=(am_1\otimes bm_2)$, where $a,b\in\alg$ and $m_i\in M_i$.
The important
observation is, for $\alg$ equipped with a bialgebra structure, the algebra morphism
$\Delta$ allows the $\alg\otimes\alg$-module $M_1\otimes M_2$ to become
a $\alg$-module, i.e. $a(m_1\otimes m_2)=a_{(1)}m_1\otimes a_{(2)}m_2$.
This procedure is applicable more than once, giving for example
a $\alg$-module structure on $M_1\otimes M_2\otimes M_3$ by
$\Delta^2(a)(m_1\otimes m_2\otimes m_3)$.
Recall that there are two possibilities for $\Delta^2$, namely
$(\mathrm{id}\otimes\Delta)\circ\Delta$ and
$(\Delta\otimes\mathrm{id})\circ\Delta$ corresponding to the representations
$M_1\otimes (M_2\otimes M_3)$, $(M_1\otimes M_2)\otimes M_3$ respectively.
For vector spaces there are the natural isomorphisms
$(V_1\otimes V_2)\otimes V_3\simeq V_1\otimes( V_2\otimes V_3)$ and
$\mathbb{K}\otimes V\simeq V\simeq V\otimes\mathbb{K}$ due to the definition of the tensor product.
In the case of modules over the algebra,
these isomorphisms must be the appropriate isomorphisms compatible with the
structure. Indeed, if $\alg$ is a bialgebra, the isomorphisms
between $(M_1\otimes M_2)\otimes M_3\simeq M_1\otimes( M_2\otimes M_3)$ and
$\mathbb{K}\otimes M\simeq M \simeq M\otimes\mathbb{K}$ are $\alg$-module morphisms, with
$\mathbb{K}$ as a $\alg$-module by the trivial action $a\lambda=\epsilon(a)\lambda$.
The existence of these properties can be expressed by saying that the category of
representations of a Hopf algebra is {\it monoidal}, because the necessary compatibility
axioms for the tensor product in a monoidal category are satisfied.
For a coassociative bialgebra, the associativity constraint $\Phi$ of the category
is trivial. If the bialgebra is a quasi bialgebra
with a Drinfel'd associator $\Omega$, the associativity constraint of $\cat$ is given by
\begin{equation}
\Phi_{U,V,W}((u\otimes v)\otimes w)=\sum \varphi_1 u\otimes(\varphi_2 v\otimes \varphi_3 w),
\end{equation}
with the notation $\Omega=\sum \varphi_1\otimes\varphi_2\otimes\varphi_3$. Therefore,
quasi-coassociativity on the bialgebra level determines a (possibly non trivial)
 associativity constraint in the category (and the strictness is lost).
\medskip

This analogy can be extended to modules over a braided bialgebra. More precicely,
for the category of $\alg$-modules there should be isomorphisms between
two objects $\Psi:U\otimes V\simeq V\otimes U$, but the two representations can be rather
unrelated. As a matter of fact, one can define a category with such a map $\Psi$
which is then a braided or quasitensor category $\cat$, i.e. a monoidal
category with a commutativity constraint seen as a functorial
isomorphism $\Psi_{V,W}:V\otimes W\to W\otimes V$ for each two objects $U,V\in\cat$.
$\Psi$ has to fulfill the two hexagon axioms.
In general, the difference between $\Psi$ and simply the transposition map $\tau$ is that one
does not assume that $\Psi\circ\Psi=\mathrm{id}$. This is why one has to think of 'braids'
rather than simple transpositions. $\Psi$ itself is normally called the braiding of the category.
\medskip

The interesting analogy between the braided category and the underlying braided bialgebra
is that $\Psi$ can be constructed from the quasi-triangular structure $\mathcal{R}$, i.e.
\begin{equation}
\Psi_{V,W}(v\otimes w)=\sum \mathcal{R}_2 w\otimes \mathcal{R}_1 v=\mathcal{R}_{21}\tau(v\otimes w),
\end{equation}
by the action of $\mathcal{R}$
(written as $\sum \mathcal{R}_1\otimes\mathcal{R}_2$) followed by the
flip $\tau$.
Therefore, if one has a braiding between
two objects of $\cat$, this corresponds to having a generator of the braid
group.

In the case of a quasi bialgebra, the Drinfel'd associator $\Omega$
makes the braiding more complicated, because one has to take care of the brackets in multiple
tensor product expressions, for example
\begin{equation}
(U\otimes V\otimes W\otimes Z)\longrightarrow (((U\otimes V)\otimes W)\otimes Z).
\end{equation}
It turns out that one can choose a unique system of parenthesis all opening on the left
(or on the right) and the action by $\Omega$ allows to change the bracket structure,
for example
$((u\otimes v)\otimes w)\rightarrow\Omega(u\otimes (v\otimes w))$, or
$((u\otimes v)\otimes (w\otimes z))\rightarrow
(\Delta\otimes\mathrm{id}\otimes\mathrm{id})[\Omega^{-1}](((u\otimes v)\otimes w)\otimes z)$.
It is not possible, to braid the last expression $((u\otimes v)\otimes (w\otimes z))$
in the 2. and 3.
argument without multiplying before with
$(\Omega\otimes\mathrm{id})(\Delta\otimes\mathrm{id}\otimes\mathrm{id})[\Omega^{-1}]$,
to enclose $v$ and $w$ into a pair of brackets.
This means that a representation of the braid group has to take this into account.
Generally, then, for any generator $\sigma_i$ of the braid
group, one can define a representation $\varrho$ by \cite{kassel}
\begin{equation}
\varrho(\sigma_i)\;(...(v_1\otimes v_2)\otimes...v_{n-1})\otimes v_n)=
\Omega^{-1}_{i}\tau_{i,i+1}\mathcal{R}_{i,i+1}(\Omega_{i}(v_1\otimes...\otimes v_n))
\end{equation}
to always ensure the right positions of the brackets. Not to be confused with the
Drinfel'd associator (but constructed from it), $\Omega_k$ is defined to be
\begin{equation}
\Omega_k=\Delta_{\mathrm{L}}^{k-2}(\Omega)\otimes\mathrm{id}^{\otimes n-k-1}
\end{equation}
with
\begin{equation} \Delta_{\mathrm{L}}(v_1\otimes...\otimes v_n)=
(\Delta(v_1)\otimes v_2\otimes...\otimes v_n),\end{equation}
and $\Omega_1$ is the identity.
This representation shifts the brackets into the proper position,
applies the braid, and puts the brackets back to place afterwards.



\newpage
\begin{thebibliography}{54}

\bibitem{magnetic}

 A.~Connes, M.~R.~Douglas and A.~S.~Schwarz,
  {\it ``Noncommutative geometry and matrix theory: Compactification on tori,''}
  JHEP {\bf 9802} (1998) 003
  [arXiv:hep-th/9711162].


  V.~Schomerus,
  {\it ``D-branes and deformation quantization,''}
  JHEP {\bf 9906} (1999) 030
  [arXiv:hep-th/9903205].

  N.~Seiberg and E.~Witten,
  {\it ``String theory and noncommutative geometry,''}
  JHEP {\bf 9909} (1999) 032
  [arXiv:hep-th/9908142].

 D.~Bigatti and L.~Susskind,
  {\it ``Magnetic fields, branes and noncommutative geometry,''}
  Phys.\ Rev.\ D {\bf 62}, 066004 (2000)
  [arXiv:hep-th/9908056].

 P.~Bouwknegt and V.~Mathai,
  {\it ``D-branes, B-fields and twisted K-theory,''}
  JHEP {\bf 0003} (2000) 007
  [arXiv:hep-th/0002023].

\bibitem{connes}

  A.~Connes,
  {\it ``Noncommutative geometry,''}, Academic Press (1994)

\bibitem{dijkgraaf}

  P.~Roche, V.~Pasquier and R.~Dijkgraaf,
  {\it ``Quasihopf Algebras, Group Cohomology And Orbifold Models,''}
  Nucl.\ Phys.\ Proc.\ Suppl.\  {\bf 18B}, 60 (1990).

\bibitem{sharpe}

E.~R.~Sharpe,
  {\it ``Discrete torsion,''}
  Phys.\ Rev.\ D {\bf 68}, 126003 (2003)
  [arXiv:hep-th/0008154].

  E.~R.~Sharpe,
  {\it ``Discrete torsion and gerbes. II,''}
  arXiv:hep-th/9909120.

  E.~R.~Sharpe,
  {\it ``Discrete torsion and gerbes. I,''}
  arXiv:hep-th/9909108.

 

\bibitem{gawedzki}

  K.~Gaw\c{e}dzki,
  {\it ``Topological actions in two-dimensional quantum field
theories,''}, in {\it "Nonperturbative quantum field theory"},
Cargese (1987)


\bibitem{kassel}
 C.~Kassel,
  {\it ``Quantum groups,''} Graduate text in mathematics,  Springer (1995)

\bibitem{chari}
V. Chari and A. Pressley, {\it ''Quantum Groups''}, Cambridge University Press (1994)

\bibitem{polychronakos}
  A.~P.~Polychronakos,
  {\it ``Abelian Chern-Simons theory in (2+1)-dimensions,''}
  Annals Phys.\  {\bf 203} (1990) 231.

\bibitem{jackiw}
  R.~Jackiw,
  {\it ``Topological Investigations Of Quantized Gauge Theories,''}
 {\it Les Houches Summer School on Theoretical Physics: Relativity
  Groups and Topology}, Les Houches (1983)


\bibitem{hitchin}

 N.~J.~Hitchin,
  {\it ``Lectures on special Lagrangian submanifolds,''}
  arXiv:math.dg/9907034.

\bibitem{sharpeMtheory}

  E.~R.~Sharpe,
  {\it ``Analogues of discrete torsion for the M-theory three-form,''}
  Phys.\ Rev.\ D {\bf 68}, 126004 (2003)
  [arXiv:hep-th/0008170].

\bibitem{seki}

  S.~Seki,
  {\it ``Discrete torsion and branes in M-theory from the mathematical view point,''}
  Nucl.\ Phys.\ B {\bf 606} (2001) 689
  [arXiv:hep-th/0103117].

\bibitem{segal}

  G.~Segal,
  {\it ``Two-dimensional conformal field theories and modular
   functors,''} in {\it  Proceedings, Mathematical physics}, Swansea
(1988)

\bibitem{newgawedzki}
 K.~Gaw\c{e}dzki and N.~Reis,
  {\it``WZW branes and gerbes,''}
  Rev.\ Math.\ Phys.\  {\bf 14} (2002) 1281
  [arXiv:hep-th/0205233].

 K.~Gawedzki,
  {\it ``Abelian and non-Abelian branes in WZW models and gerbes''},
  Commun.\ Math.\ Phys.\  {\bf 258} (2005) 23
  [arXiv:hep-th/0406072].

  K.~Gawedzki, R.~R.~Suszek and K.~Waldorf,
  {\it ``WZW orientifolds and finite group cohomology"},
  arXiv:hep-th/0701071.


\bibitem{thesis}

J.-H. ~Jureit, PhD Thesis, Kiel University (to appear)

\bibitem{simply}

K.~Gaw\c{e}dzki and N.~Reis,
  {\it ``Basic gerbe over non simply connected compact groups,''}
  arXiv:math.dg/0307010.

\bibitem{willerton}
S. Willerton, {\it ''The twisted drinfeld double of a finite group via gerbes and finite groupoids''},
[arXiv:math.QA/0503266v1]

\bibitem{coste}
  D.~Altschuler and A.~Coste,
  {\it ``Quasiquantum Groups, Knots, Three Manifolds, And Topological Field
  Theory,''}
  Commun.\ Math.\ Phys.\  {\bf 150} (1992) 83
  [arXiv:hep-th/9202047].

\bibitem{maillard}
  D.~Altschuler, A.~Coste and J.~M.~Maillard,
  {\it ``Representation theory of twisted group double''},
  Annales Fond.\ Broglie {\bf 29} (2004) 681
  [arXiv:hep-th/0309257].
 

\bibitem{sft}
H.~Hata, K.~Itoh, T.~Kugo, H.~Kunitomo and K.~Ogawa,
  {\it ``Gauge String Field Theory For Torus Compactified Closed
String,''}
  Prog.\ Theor.\ Phys.\  {\bf 77} (1987) 443.

\bibitem{schomerus}

  G.~Mack and V.~Schomerus,
  {\it ``QuasiHopf quantum symmetry in quantum theory,''}
  Nucl.\ Phys.\ B {\bf 370} (1992) 185.

\bibitem{bais}
  F.~A.~Bais, P.~van Driel and M.~de Wild Propitius,
  {\it ``Anyons In Discrete Gauge Theories With Chern-Simons Terms,''}
  Nucl.\ Phys.\ B {\bf 393}, 547 (1993)

\bibitem{braided}
  Z.~Fiedorowicz {\it "The symmetric bar construction"}, preprint
 available at {\tt http://www.math.ohio-state.edu/~fiedorow/}

\bibitem{dijkgraafwitten}
  R.~Dijkgraaf and E.~Witten,
  {\it ``Topological Gauge Theories And Group Cohomology,''}
  Commun.\ Math.\ Phys.\  {\bf 129} (1990) 393.

\bibitem{bantay}
  P.~Bantay,
  {\it ``Algebraic aspects of orbifold models,''}
  Int.\ J.\ Mod.\ Phys.\ A {\bf 9}, 1443 (1994)
  [arXiv:hep-th/9303009].

\bibitem{vafa}
 C.~Vafa,
  {\it ``Modular Invariance And Discrete Torsion On Orbifolds,''}
  Nucl.\ Phys.\ B {\bf 273}, 592 (1986).

\bibitem{douglas} M.~R.~Douglas,
  {\it``D-branes and discrete torsion,''}
  arXiv:hep-th/9807235.

\bibitem{jurco}
   P.~Aschieri   {\it ``Gerbes, M5-brane anomalies and E(8) gauge theory,''}
  JHEP {\bf 0410} (2004) 068
  [arXiv:hep-th/0409200].






\end{thebibliography}
\end{document}